\documentclass[10pt,journal]{IEEEtran}
\usepackage[latin9]{inputenc}
\usepackage{array}
\usepackage{bm}
\usepackage{multirow}
\usepackage{amsmath}
\usepackage{amssymb}
\usepackage{graphicx}
\usepackage{epsfig}
\usepackage{amsfonts}
\usepackage{cite}
\usepackage{color}
\usepackage{bm}
\usepackage{booktabs}
\usepackage{multirow}
\usepackage{comment}
\usepackage{epstopdf}
\usepackage{stfloats}
\usepackage{makecell}
\usepackage{stackengine}
\usepackage{threeparttable}
\ifCLASSOPTIONcompsoc
\usepackage[caption=false,font=normalsize,labelfon
t=sf,textfont=sf]{subfig}
\else
\usepackage[caption=false,font=footnotesize]{subfi
	g}
\fi

\newcolumntype{P}[1]{>{\centering\arraybackslash}p{#1}}

\usepackage{graphicx}
\begin{document}
\title{On the Distribution of SINR for Widely Linear MMSE MIMO Systems with Rectilinear or Quasi-Rectilinear Signals}
\author{Wei Deng, \emph{Student Member}, \emph{IEEE},
	Yili Xia, \emph{Member}, \emph{IEEE},
	Zhe Li, \emph{Member}, \emph{IEEE}, and
	Wenjiang Pei  

\thanks{This paper was presented in part at the IEEE International Symposium on Personal, Indoor and Mobile Radio Communications (PIMRC), London, U.K., Aug. 2020\cite{Li2020conf}.}
\thanks{W. Deng, Y. Xia and W. Pei are with the School of Information Science and Engineering, Southeast University, 2 Sipailou, Nanjing 210096, P. R. China. (e-mail: dengwei1@seu.edu.cn; yili\_xia@seu.edu.cn; wjpei@seu.edu.cn)}
\thanks{Z. Li is with the School of Electronic and Information Engineering, Soochow University, Suzhou 215006, China (e-mail: lizhe@suda.edu.cn)}
}

\maketitle
\begin{abstract}
Although the widely linear least  mean square error (WLMMSE) receiver has been an appealing option for multiple-input-multiple-output (MIMO) wireless systems, a statistical understanding on its pose-detection signal-to-interference-plus-noise ratio (SINR) in detail is still missing. To this end, we consider a WLMMSE MIMO transmission system with rectilinear or quasi-rectilinear (QR) signals over the uncorrelated Rayleigh fading channel and investigate the statistical properties of its SINR for an arbitrary antenna configuration with $N_t$ transmit antennas and $N_r$ receive ones. We first derive an analytic probability density function (PDF) of the SINR in terms of the confluent hypergeometric function of the second kind, for WLMMSE MIMO systems  with an arbitrary $N_r$ and $N_t=2, 3$.
For a more general case in practice, i.e., $N_t>3$, we resort to the moment generating function to obtain an approximate but closed form PDF under some mild conditions, which, as expected, is more Gaussian-like as $2N_r-N_t$ increases. The so-derived PDFs are able to provide key insights into the WLMMSE MIMO receiver in terms of the outage probability, the symbol error rate, and the diversity gain, all presented in closed form. In particular, its diversity gain and the gain improvement over the conventional LMMSE one are explicitly quantified as $N_r-(N_t-1)/2$ and $(N_t-1)/2$, respectively. Finally, Monte Carlo simulations support the analysis.
\end{abstract}
\IEEEpeerreviewmaketitle{}

\section{Introduction}

\IEEEPARstart
{T}{he} concept of widely linear (WL) processing, originally introduced in \cite{Picinbono_1995}, has received a great interest for wireless communication systems in recent years, since it brings additional performance gains over its linear counterpart when second order noncircular (improper) constellations are employed \cite{Chevalier2006,Xiao2009,Mandic_Book2,Schreier_2010,Ye2013,Bavand2018}. One particular category of such improper constellations comprises of the rectilinear and quasi-rectilinear (QR) modulations, in which the rectilinear modulation corresponds to mono-dimensional schemes, such as binary phase shift keying (BPSK), amplitude shift keying {\color{black}{and}} amplitude modulation, {\color{black}{while}} the QR modulation refers to {\color{black}{complex-valued}} examples like $\pi$/2-BPSK, offset quadrature phase shift keying (OQPSK) {\color{black}{and}} offset quadrature amplitude modulation (OQAM), which can be considered as filtered versions of a rectilinear one, after a de-rotation operation \cite{Ding1998}. Popularity of rectilinear and QR signals in communication systems can be found in \cite{Mirbagheri2006,Sterle2007,Kuchi2009,Aghaei2010,  Chevalier2011,Caus2014,Deng2019,Chevalier2014,KimYun2018,Zhe2018}, owing to the fact that they can not only preserve an extra degree of freedom compared with signals taken from a proper constellation, but also achieve high spectral efficiency at a low computational cost.

As a computationally attractive option of multiple-input-multiple-output (MIMO) receiver structures, the linear minimum mean-square error (LMMSE) receiver provides reliable data transmissions at a polynomial computational complexity and serves as building blocks of the state-of-the-art large-scale wireless systems \cite{Golden1999,Ran2017,Choi2017}. {\color{black}{However}}, for improper signals, the widely linear MMSE (WLMMSE) receiver stands as a more
efficient option, compared with the LMMSE one, since it provides as much as a two-fold performance improvement {\color{black}{at a slightly higher computational cost}}. For this reason, the WLMMSE receiver has been extensively used in numerous applications where improper signals appear due to their underlying generating
physics \cite{Sterle2007,Kuchi2009,Aghaei2010,Chevalier2011,Chevalier2014,Caus2014,Deng2019}. For instance, the WLMMSE receiver in \cite{Chevalier2011} jointly exploits both the real-valued nature of the source symbols and the space-time structure of the Alamouti scheme to improve interference cancellation. In the context of multi-antenna radio frequency identification (RFID) systems, the enhanced anti-collision performance of the WLMMSE tag signal recovery method has been verified by real-world experiments \cite{Deng2019}. Moreover, in MIMO systems with the OQAM orthogonal frequency division multiplexing waveform, a candidate for beyond 5G and future Internet of things \cite{Chen2018}, the optimization of precoding and decoding matrices is built upon the WLMMSE criterion to exploit the improper nature of OQAM symbols \cite{Caus2014}.

To strike a balance between the complexity and the performance of a MIMO receiver in practice, one of the most challenging tasks lies in characterizing its output signal-to-interference-plus-noise ratio (SINR) behavior. In the literature, although a lot of efforts have been made to analyze the statistical distribution of the SINR for LMMSE receivers in last two decades \cite{Poor1997,Tse1999,Li2006,Kim2008}, it is very recently that the general probability density function (PDF) of the output SINR for arbitrary antenna configurations has been established \cite{Lim2019}. However, the SINR distribution analysis for WLMMSE MIMO systems is even more challenging, since a complete statistical description of the output SINR requires a joint processing on both the covariance matrix and the pseudo-covariance matrix of the improper transmit signals, the latter of which vanishes in the above mentioned LMMSE models \cite{Poor1997,Tse1999,Li2006,Kim2008,Lim2019}.
In \cite{Schober2004}, an adaptive WLMMSE detector has been developed in code division multiple access systems, and the analytic output SINR at steady state highlights the performance gain of the WL processing framework over its strictly linear counterpart. In \cite{Kuchi2009}, a WLMMSE demodulator was implemented in MIMO systems, and its SINR  was {\color{black}{expressed}} in a quadratic form, based on which an upper bound and an asymptotic evaluation on the average symbol error rate (SER) have been made possible.  However, neither of them attempts to address the general SINR distribution of the WLMMSE receiver. To fill this void, we consider a WLMMSE MIMO system with rectilinear or QR transmit signals, and investigate the statistical properties of its SINR in detail. The main contributions of this paper are summarized as follows:

\begin{itemize}
\item Based on the preliminary results in \cite{Li2020conf}, the general PDF of the SINR for a WLMMSE MIMO receiver with $N_t$ transmit antennas and $N_r$ receive antennas is derived, which involves an ($N_t\!-\!1$)-fold integral with the confluent hypergeometric function in the integrand. By doing so, mathematically tractable PDFs are deduced for an arbitrary $N_r$ and $N_t=2, 3$, whose analyticity is rigorously proved.

	\item For the more general case, i.e., $N_t>3$, we alternatively resort to the moment generating function (MGF), and omit higher order terms of its Taylor-series expansion, so as to yield an approximate PDF of the SINR. The resulting PDF is of a concise and closed form, which is valid for {\color{black}{both}} small-scale MIMO systems under rather mild conditions {\color{black}{and}} massive MIMO systems. Particularly, it becomes more Gaussian-like as $2N_r-N_t$ increases.

	\item We make progress on the approximate PDF to elaborate its usefulness in typical MIMO system performance metrics, including outage probability, SER, and diversity gain. These closed form evaluations explain the simulated and experimental results in previous studies of WLMMSE estimators \cite{Kuchi2009,Deng2019,YangLamare2015}, and provide design guidelines for engineers in practice. In particular, the diversity gain of the WLMMSE MIMO receiver and its gain improvement against the conventional LMMSE one are explicitly quantified as $N_r-(N_t-1)/2$ and $(N_t-1)/2$, respectively.
\end{itemize}

The remainder of this paper is organized as follows. In Section II, we briefly review the MIMO system model and then introduce the WLMMSE detector, as well as {\color{black}{the statistical properties of}} its SINR.
In Section III, we evaluate the SINR of WLMMSE MIMO systems in terms of its general and analytic PDFs.
In Section IV, an approximate PDF of the SINR is further derived, whose usefulness for analyzing the performance metrics, including outage probability, SER, diversity gain, of our considered MIMO system is highlighted. In Section V, the theoretical findings are verified through Monte Carlo simulations. Finally, Section VI concludes the paper.

The notations used in this paper are as follows:
\begin{itemize}
	\item Lowercase letters denote scalars, $a$, boldface letters column vectors, $\textbf{a}$, and boldface uppercase letters matrices, $\textbf{A}$. An $N \times N$ identity matrix is denoted by $\textbf{I}_N$. The superscripts $(\cdot)^{*}$, $(\cdot)^{\mathcal{T}}$, $(\cdot)^{\mathcal{H}}$ and $(\cdot)^{-1}$ denote respectively the complex conjugation, transpose, Hermitian transpose and matrix inversion operators.
	The statistical expectation operator is denoted by $E[\cdot]$, while the operator ${\rm Tr}\{\cdot\}$ return the trace of a square matrix.
	The operators $\left\Vert \textbf{a} \right\Vert$ and $\left\Vert \textbf{A} \right\Vert$ return respectively the Euclidean norm of the vector $\textbf{a}$ and the Frobenius norm of the matrix $\textbf{A}$, and $\jmath=\sqrt{-1}$.

	\item  ${\mathcal{N}}(\textbf{a},{\mathbf A})$ represents a real-valued Gaussian distribution with mean $\textbf{a}$ and covariance ${\mathbf A}$, while ${\mathcal{CN}}(\textbf{b},{\mathbf B})$ denotes a complex-valued circularly-symmetric Gaussian distribution with mean $\textbf{b}$ and covariance ${\mathbf B}$. ${\rm Gamma}(\alpha,\beta)$ represents a gamma distribution with a shape parameter $\alpha$ and a scale parameter $\beta$.
	\item $\Gamma(a)=\int\limits_0^\infty t^{a-1}e^{-t}dt$, $\gamma(a,b) = \int\limits_0^b t^{a - 1}e^{-t}dt$ and ${\Gamma _b}(a) = {\pi ^{b(b-1)/4}}\prod\limits_{i = 1}^b {\Gamma \big(a +(1-i)/2\big)}$ represent respectively the gamma function, the incomplete gamma function and the multivariate gamma function. The rising factorial operation is denoted by $(a)_b=\prod\limits_{i=0}^{b-1}(a-i)$.  The Q-function is represented by ${\mathcal Q}(x) = \frac{1}{{\sqrt {2\pi } }}\int\limits_x^\infty  {{e^{ - \frac{{{t^2}}}{2}}}dt}$, where $x \ge 0$, and it computes the area under the tail of the standard Gaussian distribution. 
\end{itemize}

\section{Preliminaries}
\subsection{System Model }
Let us consider a single point-to-point MIMO system with $N_t$ transmit and $N_r$ receive antennas, where $N_t \leq N_r $. Then, the $l$th received symbol vector, ${\bf{z}}(l) \in \mathbb{C}^{N_r \times 1}$, can be written as
\begin{gather}
{\bf{z}}(l) = {\bf{H}}{\bf{d}}(l)+{\bf{v}}(l),
\label{Eq:SYS_MODEL}
\end{gather}
where ${\bf{H}} \triangleq [{{\bf{h}}_1},{{\bf{h}}_2},\cdots,{\bf{h}}_{N_t}] \in \mathbb{C}^{N_r \times N_t}$ is a channel gain matrix, and each column vector ${\bf{h}}_j$ contains the channels from the $j$th transmit antenna to all receive antennas, where $1 \leq j \leq N_t$. The vector ${\bf{d}}(l)\triangleq[d_1(l),d_2(l),...,d_{N_t}(l)]^{\mathcal H} \in \mathbb{C}^{N_t \times 1}$ is the $l$th transmitted symbol compliant with a rectilinear or QR modulation, and ${\bf{v}}(l) \in \mathbb{C}^{N_r \times 1}$ is the complex-valued noise vector subject to ${\mathcal{CN}}({\bf 0},\sigma^2{\mathbf I}_{N_r})$. We assume each element of the channel gain matrix ${\bf{H}}$, $h_{i,j}$, conforms to independent identically distributed (i.i.d.) standard complex-valued Gaussian distribution, i.e., $h_{i,j}\sim{\mathcal{CN}}(0,1)$, for $1 \leq i \leq N_r$ and $1 \leq j \leq N_t$.  The transmit power $E_s$ is assumed to  be distributed equally over the transmit antennas.

\begin{figure}[t!]
	\centering
	\includegraphics[width=0.95\linewidth,height=0.20\linewidth]{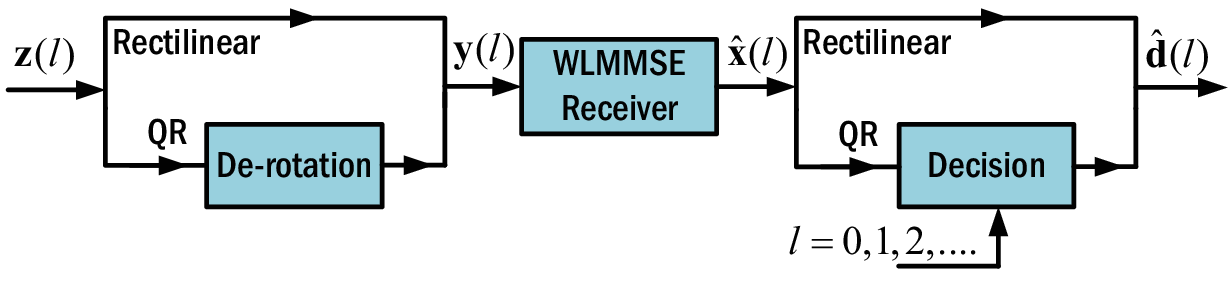}
	\caption{Flowchart of the detection process at the receiver{\color{black}\cite{chevalier2021widely}}.}
	\label{Fig:SYSTEM_MODEL2}
\end{figure}

The flowchart of the detection process at the receiver is depicted in Fig. \ref{Fig:SYSTEM_MODEL2}, where the transmitted rectilinear or QR symbol vector ${{\bf{d}}(l)}$ is estimated by the WLMMSE receiver based on the observation ${{\bf{z}}(l)}$.  According to \cite{Akbar2020}, when the $l$th transmitted symbol on the $j$th transmit antenna, ${d_j}(l)$, is QR modulated, we have
\begin{align}
{d_j}\left( l \right) = \left\{ {\begin{array}{*{20}{c}}
		{{x_j}(l) \;\;\;\;\;\;\;\;\;l\rm{\;is\;even}}\\
		{\jmath{x_j}(l) \;\;\;\;\;\;\;l\rm{\;is\;odd}},
\end{array}} \right.
\label{Eq:QR_SIG_EXPRESSION}
\end{align}
where $x_j(l)$ is an i.i.d. two-point random variable with probability ${\rm Pr}\left\{x_j(l)=-\sqrt{{E_s}/{N_t}}\right\}={\rm Pr}\left\{x_j(l)=\sqrt{{E_s}/{N_t}}\right\}=0.5$.
An inspection on \eqref{Eq:QR_SIG_EXPRESSION} indicates the symbol ${d_j}(l)$ with an odd index $l$ can be converted back to $x_j(l)$ by multiplying a de-rotation coefficient $(-\jmath)^l$. Therefore, for QR transmit signals, a de-rotation operation and a decision-making one are respectively placed before and after the WLMMSE receiver, in which the input signal is respectively multiplied by a de-rotation coefficient $(-\jmath)^l$ and a recovering coefficient $(\jmath)^l$. In this way, for both rectilinear and QR symbols, the system model in \eqref{Eq:SYS_MODEL} becomes \cite{Chevalier2018}
%

\begin{gather}
{{\bf{y}}(l)} =  {{\bf{H}}{\bf x}(l)} + {{\bf{n}}}(l),
\label{Eq:SYS_MODEL2}
\end{gather}
where ${\bf{y}}(l) = (-\jmath)^l {\bf{z}}(l) \in \mathbb{C}^{N_r \times 1}$ is the equivalent observation vector, ${\bf{x}}(l)\triangleq[x_1(l),x_2(l),...,x_{N_t}(l)]^{\mathcal T} \in \mathbb{R}^{N_t \times 1}$  is the equivalent rectilinear transmit vector, and ${\bf{n}}(l)= ({-\jmath})^l{\bf{v}}(l)$ is the equivalent noise vector, which still subjects to ${\mathcal{CN}}({\bf 0},\sigma^2{\mathbf I}_{N_r})$, according to \cite{andersen1995multivariate}. Given the detection process depicted in Fig. \ref{Fig:SYSTEM_MODEL2}, the WLMMSE detection for the QR transmit signal is inherently identical to its counterpart for the rectilinear one \cite{Chevalier2018,chevalier2021widely}. For this reason, in what follows, the rectilinear signal is adopted for analysis, because there is no need to differentiate it from the QR one. In addition, {\color{black}{whenever it is clear from the context, the symbol index $l$ has been omitted for mathematical elegance.}}
%

%
%


\subsection{WLMMSE Detector}
In order to estimate the transmitted symbol $x_j$ on the $j$th spatial stream from a complex-valued observation vector $\bf{y}$, the WL processing framework is adopted, based on both $\bf{y}$ and its conjugate counterpart $\bf{y}^*$, to give
%
%
%
\begin{gather}
{\hat{x}_j} = \bar{{\bf{w}}}_j^{{\mathcal H}}\bar{\bf{y}},
\label{Eq:WL_EST_EXPR1a}
\end{gather}
where $\bar{\bf{w}}_j \triangleq [ {{\bf{w}}_{j,1}^{{\mathcal T}}} ,{{\bf{w}}_{j,2}^{\mathcal T}} ]^{\mathcal T} \in \mathbb{C}^{2N_r \times 1}$ and ${\bar{{\bf y}}}\triangleq[{\bf{y}}^{\mathcal T}, {\bf{y}}^{\mathcal H}]^{\mathcal T} \in \mathbb{C}^{2N_r \times 1}$ are respectively the augmented weight coefficient vector and the augmented observation vector. As shown in Fig. \ref{Fig:SYSTEM_MODEL2}, by applying the WLMMSE criterion, that is, minimizing the estimation error $e_j$ in the mean square sense, where $e_j \triangleq x_j - {\hat{x}_j}$, the yet-to-be-detected symbol on the $j$th spatial stream, ${\hat{x}_j}$, can be obtained as \cite{Picinbono_1995}
%
%
\begin{align}
{\hat{x}_j} =  \bar{{{\bf{h}}}}_j^{\mathcal H}\left({\bar{\bf{{T}}}} + {\rho ^{-1}}{\mathbf{I}_{2N_r}}\right)^{-1} \bar{\bf{y}}, \label{Eq:WL_EST_EXPR3a}
\end{align}
where ${\bar{\bf{h}}_j}\triangleq[{{\bf{h}}_j^{\mathcal T}},{{\bf{h}}_j^{\mathcal H}}]^{\mathcal T} \in \mathbb{C}^{2N_r \times 1}$  is the $j$th augmented channel column vector, and the Hermitian matrix $\bar{\bf{T}} \in \mathbb{C}^{2N_r \times N_t}$ is given by
\[{\bar{\bf{{T}}}} = \left[ {\begin{array}{*{20}{c}}
{\bf H}{\bf H}^{\mathcal H}&{\bf H}{\bf H}^{\mathcal T} \\
{\bf H}^{*}{\bf H}^{\mathcal H}&{\bf H}^{*}{\bf H}^{\mathcal T}
\end{array}} \right].\]
The coefficient $\rho$ in \eqref{Eq:WL_EST_EXPR3a} is defined as $\rho \triangleq E_s/(\sigma^2N_t)$, which evaluates the average signal-to-noise ratio (SNR) per symbol per antenna.

\subsection{SINR of the WLMMSE Detector}

The WLMMSE estimate ${\hat{x}_j}$ in \eqref{Eq:WL_EST_EXPR3a} can be expressed as
\begin{align}
{\hat{x}_j} = {\epsilon_j}{x_j} + {\xi _j},\label{Eq:WL_EST_EXPR3b}
\end{align}
where the effective channel gain ${\epsilon_j}$ and the interference-plus-noise component ${\xi _j}$ are respectively defined as{\color{black}\cite[Eqs. (8), (9)]{Li2020conf}}
\begin{align}
{\epsilon_j}&\triangleq \bar{{{\bf{h}}}}_j^{\mathcal H}    ({\bar{\bf{{T}}}} + {\rho^{-1}}{\mathbf{I}_{2N_r}})^{-1}{\bar{\bf{h}}_j}\label{Eq:epsilon_j},\\
{\xi _j} &\triangleq  \bar{{{\bf{h}}}}_j^{\mathcal H} ({\bar{\bf{{T}}}} + {\rho^{-1}}{\mathbf{I}_{2N_r}})^{-1}(\bar{\bf{y}} -  {x_j} {\bar{\bf{h}}_j} ). \label{Eq:xi_j}
\end{align}
%

\textit{Lemma 1:} Let
%
${\tau_j} \triangleq  {E\left[ {{{\left| {{\epsilon_j}{x_j}} \right|}^2}} \right]}/{E\left[ {{{\left| {{\xi _j}} \right|}^2}} \right]}$
%
denote the SINR of the WLMMSE detector on the $j$th spatial stream. {\color{black}{By taking the statistical expectation}} with respect to the signal in \eqref{Eq:epsilon_j} and  \eqref{Eq:xi_j}, {\color{black}we have \cite[Eq. (20)]{Li2020conf}}
%
\begin{align}
{\tau_j} \!=\! \sum\limits_{k = 1}^{{N_t} \!-\! 1} {\frac{1}{{{\lambda _k} \!+\! {1}/(2\rho)}}} \widetilde h_{k,j}^2  + 2\rho\sum\limits_{k = {N_t}}^{2{N_r}} {\widetilde h_{k,j}^2},
\label{Eq:WL_SINR_EXPR3}
\end{align}
where $\lambda_{1}>\lambda_{2}>\cdots>\lambda_{N_t-1}>0$ are the positive eigenvalues of the real-valued Wishart matrix ${\mathcal{W}}_{N_t-1}(\frac{1}{2}{\bf I}_{N_t-1},2N_r)$, and the i.i.d. random variable (RV) $\widetilde h_{k,j}\sim{\mathcal{N}}(0,1/2)$.

\textit{Proof:} See (10)-(23) in \cite{Li2020conf}. $\hfill\blacksquare$

\textit{Remark 1:} Observe that on the right hand side (RHS) of \eqref{Eq:WL_SINR_EXPR3}, the term, ${\frac{1}{{{\lambda _k} \!+\! {1}/(2\rho)}}} \widetilde h_{k,j}^2$, is a gamma RV with the shape parameter $1/2$ and the scale parameter ${2/( {2{\lambda _k} \!+\!  {\rho^{-1}}})}$, according to Appendix \ref{App:GammaRV}, while the summation, $2\rho\sum\limits_{k = {N_t}}^{2{N_r}} {\widetilde h_{k,j}^2}$, can be viewed as a single gamma RV with the shape parameter ${(2{N_r}\!-\!N_t+1)/2}$, and the scale parameter $2\rho$. In this way, the SINR $\tau_j$ can be regarded as a summation of $N_t$ gamma RVs, in which the shape parameter $ \alpha_k $ and the scale parameter $\beta_k$ of the $k$th gamma RV $Z_k \sim {\rm Gamma}(\alpha_k,\,\beta_k) $ are respectively given by
\begin{align}\label{Eq:WL_SINR_PARA_EXPR1a}
{\alpha _k}  &= \left\{ {\begin{array}{*{20}{c}}
	{1/2}&{1 \!\leqslant\! k  \!\leqslant\! {N_t}\!-\!1}  \\
	{(2{N_r}\!-\!N_t+1)/2}  &  {k \!=\! N_t }
	\end{array}} \right.,\, \\
{\beta _k}  &=  \left\{ {\begin{array}{*{20}{c}}
	{2/( {2{\lambda _k} \!+\!  {\rho^{-1}}  }    )} &{1 \!\leqslant \!k\!  \leqslant\! {N_t}-1} \\
	{2\rho} &{k \!=\! N_t }
	\end{array}}. \right.
\label{Eq:WL_SINR_PARA_EXPR1}
\end{align}
%

\textit{Remark 2:} The SINR $\tau_j$ of the WLMMSE detector in \eqref{Eq:WL_SINR_EXPR3} is different from that of the LMMSE one, which has been proven to be the sum of $N_t-1$ exponential RVs plus a chi-square RV \cite[Eq. (10)]{Kim2008}, \cite[Eq. (12)]{Lim2019}. Especially, when $N_t=1$, an exact 3 dB SINR gain can be achieved for a MIMO system with an arbitrary $N_r$ by using the WLMMSE detector.

\section{Analytic SINR Distribution Analysis}
We next investigate the analytic SINR distribution for the considered WLMMSE MIMO system. By exploiting the results of \textit{Lemma 1}, the general PDF of the output SINR is presented by the following theorem.

\textit{Theorem 1:} Given the WLMMSE receiver in \eqref{Eq:WL_EST_EXPR3a}, the general PDF of the SINR ${\tau_j}$, denoted by $f(\tau_j)$, can be expressed as
\begin{align}
f(\tau_j) &= \int\limits_0^{\infty}   \int\limits_{0}^{\lambda_1} \cdots \!\!\!\int\limits_{0}^{\lambda_{N_t-2}}  f({\tau_j}|{\lambda _1},{\lambda _2},...,{\lambda _{{N_t} - 1}})  \nonumber \\
&\times f_{\bf \Lambda}({\lambda _1},{\lambda _2},\ldots,{\lambda_{N_t-1}}) d\lambda_{N_t-1}\cdots d\lambda_2d\lambda_1,
\label{Eq:WL_SINR_EXPR1}
\end{align}
where $f({\tau_j}|{\lambda _1},{\lambda _2},...,{\lambda _{{N_t} - 1}})$ is the joint conditional PDF of the sum of $N_t$ gamma RVs with parameters described by \eqref{Eq:WL_SINR_PARA_EXPR1a} and \eqref{Eq:WL_SINR_PARA_EXPR1}. It can be further expanded as
\begin{align}
&\!f({\tau_j}|{\lambda _1},{\lambda _2},...,{\lambda _{{N_t} - 1}}) \!=\! \frac{{{\tau_j}^{{N_r} - 1}{e^{ - \frac{{{\tau_j}}}{{{\beta _1}}}}}}}{{\prod\limits_{k = 1}^{{N_t}} {\beta_k^{\alpha _k} \Gamma(N_r)} }} \nonumber\\
&\!\!\times \!\!\Phi _2^{(N_t\!-\!1)}\!\!\left(\!\! {{\alpha _2},{\alpha _3},\!...,{\alpha _{N_t}};{N_r};\left( \frac{1}{\beta_1}\!\!-\!\!\frac{1}{\beta_2}\right){\tau_j},\!...,\left( \frac{1}{\beta_1}\!\!-\!\!\frac{1}{\beta_{N_t}}\right){\tau_j}} \!\!\right)\!\!,
\label{Eq:WL_SINR_PDF_EXPR1}
\end{align}
where $\Phi _2^{(n)}(\cdot;\cdot;\cdot)$ is the multivariate confluent hypergeometric function, defined as \cite[pp.162-163]{mathai1978h}
\begin{align}
&\Phi _2^{(n)}\!\left( {{b_1},{b_2},...,{b_n};c;{x_1},{x_2},...,{x_n}} \right) \nonumber \\
&\triangleq \sum\limits_{{i_1} = 0}^\infty  {\mathop \sum \limits_{{i_2} = 0}^\infty \dots {\sum\limits_{i_n = 0}^\infty  \frac{{\left({b_1} \right)}_{i_1}{\left({b_2} \right)}_{i_2}...{{\left( {{b_n}} \right)}_{{i_n}}}}{{(c )}_{i_1 + i_2+... + i_n}} \frac{x_1^{i_1}}{i_1!}\frac{x_2^{i_2}}{i_2!}\cdots \frac{x_n^{{i_n}}}{i_n!} } }.
\label{Eq:WL_SINR_PDF_SUB_EXPR1}
\end{align}
The function $f_{\bf \Lambda}({\lambda _1},{\lambda _2},\ldots,{\lambda_{N_t-1}})$ is the joint PDF of the positive eigenvalues of the real-valued Wishart matrix ${\mathcal{W}}_{N_t-1}(\frac{1}{2}{\bf I}_{N_t-1},2N_r)$, given by
\begin{align}
f_{\bf \Lambda}&({\lambda _1},{\lambda _2},\ldots,{\lambda_{N_t-1}}) \!=\! \frac{\pi ^{(N_t-1)^2/2}}{{\Gamma _{N_t-1}( N_r){\Gamma _{N_t-1}}\left( {\frac{N_t-1}{2}} \right)}}    \nonumber\\
&\times \prod\limits_{k = 1}^{N_t-1} {e^{-\lambda _k}} (\lambda_k)^{N_r-\frac{N_t}{2}}\prod\limits_{k < l}^{N_t-1} {\left( {{\lambda _k} \!-\! {\lambda _l}} \right)}.
\label{Eq:REAL_WISHART_DIST}
\end{align}

\textit{Proof:}
According to \textit{Lemma 1}, ${\tau_j}$ is a linear combination of $N_t$ gamma RVs with different {\color{black}{shape and scale}} parameters given by \eqref{Eq:WL_SINR_PARA_EXPR1a} and \eqref{Eq:WL_SINR_PARA_EXPR1}. Therefore, upon substituting \eqref{Eq:WL_SINR_PARA_EXPR1a} and  \eqref{Eq:WL_SINR_PARA_EXPR1} into \eqref{Eq:PDF_OF_GAMMA_WITH_DIFF_PARAS} in Appendix \ref{App:SumofGammaRVwithDiffParas}, we obtain the joint conditional PDF $f({\tau_j}|{\lambda _1},{\lambda _2},...,{\lambda _{{N_t} - 1}})$ in \eqref{Eq:WL_SINR_PDF_EXPR1}. On the other hand,  based on \eqref{Eq:realwisharteigpdf} in  Appendix \ref{App:realwishart}, the joint PDF of the ordered positive eigenvalues of the real-valued Wishart matrix ${\mathcal{W}}_{N_t-1}(\frac{1}{2}{\bf I}_{N_t-1},2N_r)$, denoted by $f_{\bf \Lambda}({\lambda _1},{\lambda _2},\ldots,{\lambda_{N_t-1}})$, can be described by \eqref{Eq:REAL_WISHART_DIST}. Then, with \eqref{Eq:WL_SINR_PDF_EXPR1} and \eqref{Eq:REAL_WISHART_DIST}, the general PDF $f{\tau_j}$ in \eqref{Eq:WL_SINR_EXPR1} is proved by marginalization. $\hfill\blacksquare$


\textit{Remark 3:} As stated in \textit{Lemma 1}, $\lambda_{1}, \lambda_{2}, \ldots, \lambda_{N_t-1}$ are ordered eigenvalues of a real-valued Wishart matrix ${\mathcal{W}}_{N_t-1}(\frac{1}{2}{\bf I}_{N_t-1},2N_r)$, therefore their joint distribution in \eqref{Eq:REAL_WISHART_DIST} is different from that of a complex-valued Wishart matrix, {\color{black}{typically involved in}} the SINR distribution analysis of LMMSE receivers \cite{Kim2008,Lim2019}.

%
%
\textit{Remark 4:} The algebraic computation on the RHS of \eqref{Eq:WL_SINR_EXPR1} involves an $(N_t-1)$-fold integral, and the integrand $f({\tau_j}|{\lambda _1},{\lambda _2},...,{\lambda _{{N_t} - 1}})$ in \eqref{Eq:WL_SINR_PDF_EXPR1} contains the multivariate confluent hypergeometric function $\Phi _2^{(n)}(\cdot;\cdot;\cdot)$ which prohibits a simple integral representation \cite{Aalo2005,Ansari2017}. In general, the PDF $f(\tau_j)$ is mathematically intractable even with some advanced integration computing tools, such as Matlab and Mathematica. However, for a small $N_t$, i.e., $N_t=2$ or $N_t=3$, $\Phi _2^{(n)}(\cdot;\cdot;\cdot)$ reduces to the confluent hypergeometric function of the first kind \cite{olver2010nist}, which gives the following two corollaries.

\textit{Corollary 1.1:} Given the WLMMSE receiver in \eqref{Eq:WL_EST_EXPR3a}, the analytic PDF of the SINR $\tau_j$ for the $N_r \times 2 $ MIMO system, denoted by ${f_{{N_r} \!\times\! 2}}(\tau_j)$, can be expressed as
%
\begin{align}
&{f_{{N_r} \!\times\! 2}}(\tau_j) \!\!=\!\! \frac{  {\tau_j ^{{N_r} - 1}}  {e^{-\frac{\tau_j}{2\rho}}} } {{(2\rho  )^{2N_r}}(\Gamma(N_r))^2}
 \nonumber \\
&\!\!\times \!\! \sum\limits_{m = 0}^\infty  \!\frac{{(N_r\!-\!\frac{1}{2})}_m\tau_j^m\Gamma(N_r\!+\!m)}{{{(N_r)}_m}2^m\rho^m m!}U\!\!\left(\!\!N_r\!+\! m,N_r\!+\! m\!+\!\frac{3}{2} ,\frac{\tau_j \!+\!1}{2\rho}\!\!\right)\!\!,
\label{Eq:WL_SINR_PDF_Nrx2_5}
\end{align}
%
where $U(\cdot, \cdot, \cdot)$ is the confluent hypergeometric function of the second kind \cite{olver2010nist}.
\newcounter{TempEqCnt}
\setcounter{TempEqCnt}{\value{equation}}
\begin{figure*}[hb]
	\hrulefill
	\setcounter{equation}{23}
	\begin{align}
		{f_{{N_r} \!\times\! 3}}(\tau_j) &\!=\! \frac{ {e^{-\frac{\tau_j}{2\rho}}}  }{{\sqrt{2\rho} }}\sum\limits_{{m_1} = 0}^\infty  {\mathop \sum \limits_{{m_2} = 0}^\infty  \sum\limits_{m_3 = 0}^\infty  {\sum\limits_{m_4 = 0}^\infty  {\frac{{{(-1)^{m_3+m_4}(\frac{1}{2})_{m_1}(-\frac{1}{2})_{m_4}}{{(N_r\!\!-\!\! 1)}_{m_2}}{\left({N_r\!+\!m_3\!\!-\!\!\frac{1}{2}} \right)}_{m_4}{\tau_j^{m_1\!+\! m_2}}}}{\sqrt 2 {\left( {{N_r}} \right)}_{{m_1} + {m_2}}{\left(N_r\!+\!m_1\!+\!m_3\!+\!\frac{3}{2} \right)}_{m_4}\sum\limits_{l = 1}^4({m_l}!)}} } }  \nonumber\\
		&\times\frac{\Gamma(m_1\!\!+\!\!2)\Gamma \left( {N_r \!+\! m_3\!\! -\!\! \frac{1}{2}} \right){\Gamma \left( {2{N_r} \!\!+  \sum\limits_{l = 1}^4 {{m_l}}    } \right)}}{\Gamma \left( {{N_r} + {m_1} + m_3 + \frac{3}{2}} \right)(2\rho) ^{2{N_r} \!+ \sum\limits_{i = 1}^3 {{m_i}}  }}U\!\!\left( {2{N_r} \!\!+\sum\limits_{l = 1}^4 {{m_l}} ,\!2{N_r} \!\!+\sum\limits_{l = 1}^4 {{m_l}}  +\!\! \frac{3}{2},\!\frac{\tau_j\!+\!1 }{2\rho}} \right).
		\label{Eq:WL_SINR_PDF_Nrx3}
	\end{align}
\end{figure*}

\textit{Proof:}
For the $N_r \times 2$ MIMO system, there exists only one single positive eigenvalue $\lambda$, and the Wishart matrix ${\mathcal{W}}_{N_t-1}(\frac{1}{2}{\bf I}_{N_t-1},2N_r)$ reduces to a gamma RV, whose PDF $f_{\bf \Lambda}(\lambda )$ can be deduced from \eqref{Eq:REAL_WISHART_DIST} as
\begin{gather}\label{Eq:lambdaPDF}
\setcounter{equation}{\value{TempEqCnt}}
f_{\bf \Lambda}(\lambda ) = \frac{1}{\Gamma(N_r)}{\lambda ^{{N_r} - 1}}{e^{ - \lambda}}.
\end{gather}
On the other hand, the conditional PDF of ${\tau_j}$ in \eqref{Eq:WL_SINR_PDF_EXPR1} becomes
\begin{align}\label{Eq:WL_SINR_PDF_Nrx2_1}
\!\!\!f_{N_r \times 2}({\tau_j}|\lambda) &=\! \frac{{{\tau_j}^{{N_r} - 1}{e^{ - \frac{{{\tau_j}}}{{{\beta _1}}}}}}}{ {\beta_1^{\alpha_1}\beta_2^{\alpha_2}\Gamma(N_r)} }\Phi _2^{(1)}\!\!\left(\!\! {{\alpha _2};{N_r};\left( \frac{1}{\beta_1}\!\!-\!\!\frac{1}{\beta_2} \right){\tau_j}}\!\! \right) \nonumber \\
&=\!\frac{{{\tau_j}^{{N_r} - 1}{e^{ - \frac{{{\tau_j}}}{{{\beta _1}}}}}}}{ {\beta_1^{\alpha_1}\beta_2^{\alpha_2}\Gamma(N_r)} }M\!\!\left(\!\! {{\alpha _2},{N_r},\left( \frac{1}{\beta_1}\!\!-\!\!\frac{1}{\beta_2} \right){\tau_j}}\!\! \right)\!\!,
\end{align}
where $M(\cdot,\cdot,\cdot)$ is the confluent hypergeometric function of the first kind \cite{olver2010nist}. Upon using \eqref{Eq:WL_SINR_PARA_EXPR1} to interpret the shape parameters $\alpha_1,\alpha_2$, and the scale parameters $\beta_1,\beta_2$, from \eqref{Eq:WL_SINR_PDF_Nrx2_1}, we have
\begin{align}
{f_{{N_r} \times 2}}(\tau_j | \lambda ) &= \frac{{{\tau_j ^{N_r-1}}}{e^{ -\frac{2\lambda+{\rho^{-1}}}{2} \tau_j }}\sqrt{2\lambda+\rho^{-1}}}{{\Gamma(N_r){2^{N_r}}{{\rho}^{N_r-\frac{1}{2}}}}} \nonumber \\
&\times   M\left(N_r\!-\!\frac{1}{2},{N_r},\lambda \tau_j\right).
\label{Eq:WL_SINR_PDF_Nrx2_2}
\end{align}
Now, based on \eqref{Eq:lambdaPDF} and \eqref{Eq:WL_SINR_PDF_Nrx2_2}, the PDF of $\tau_j$ for the $N_r \times 2 $ WLMMSE MIMO system can be computed by marginalization as
\begin{align}
&{f_{{N_r}\!\times\! 2}}(\tau_j)   = \frac{\tau_j ^{N_r-1}e^{-\frac{\tau_j}{2\rho}}}{2^{N_r}\left(\Gamma(N_r)\right)^2{\rho^{N_r-\frac{1}{2}}}} \nonumber\\
&\!\!\times\!\!\! \int\limits_0^\infty\!\! \sqrt{2\lambda\!+\!\rho^{-1}}{\lambda ^{N_r-1}}e^{- (1+\tau_j)\lambda }M\!\left(\!N_r\!-\!\frac{1}{2},\!{N_r},\lambda\tau_j \!\right)\!d\lambda.
\label{Eq:WL_SINR_PDF_Nrx2_3}
\end{align}

According to \cite[Eq. (13.2.2)]{olver2010nist}, the confluent hypergeometric function of the first kind, that is, $M\left(N_r\!-\!\frac{1}{2},{N_r},\lambda\tau_j \right)$ in \eqref{Eq:WL_SINR_PDF_Nrx2_3}, allows the following infinite-series expansion
\begin{equation}\label{Eq:Mexpansion}
M\left(N_r\!-\!\frac{1}{2},{N_r},\lambda\tau_j \right)=\sum\limits_{m = 0}^\infty  \frac{{(N_r\!-\!\frac{1}{2})}_m (\lambda\tau_j)^m}{{{(N_r)}_m}m!}.
\end{equation}

Then, a substitution of \eqref{Eq:Mexpansion} into \eqref{Eq:WL_SINR_PDF_Nrx2_3} gives
\begin{align}
&{f_{{N_r} \!\times\! 2}}(\tau_j) \!=\! \frac{\tau_j ^{N_r-1}e^{-\frac{\tau_j}{2\rho}}}{2^{N_r}(\Gamma(N_r))^2{\rho^{N_r-\frac{1}{2}}}}   \nonumber \\
&\!\!\times\!\!\sum\limits_{m = 0}^\infty  \frac{{(N_r\!-\!\frac{1}{2})}_m\tau_j^m}{{{(N_r)}_m}m!}\underbrace{\int\limits_0^\infty  \sqrt{2\lambda\!+\!\rho^{-1}}{\lambda ^{N_r+m-1}}e^{- (1+\tau_j)\lambda}d\lambda}_{g(m)}.
\label{Eq:WL_SINR_PDF_Nrx2_4}
\end{align}

Observe that the integral term in \eqref{Eq:WL_SINR_PDF_Nrx2_4}, $g(m)$, is in fact the Mellin transform of the term, $\sqrt{2\lambda\!+\!\rho^{-1}}{\lambda ^{N_r}}e^{-(1+\tau_j)\lambda}$, so that according to \cite{Prudnikov1990vol3} and after some algebraic manipulations, the function $g(m)$ becomes
\begin{equation}\label{Eq:gm}
g(m) \!\!=\!\! \frac{\Gamma(N_r+m)}{2^{N_r+m}\rho^{N_r+m+\frac{1}{2}}}U\!\!\left(\!\!N_r\!+\! m,\!N_r\!+\! m\!+\!\frac{3}{2},\!\frac{\tau_j \!+\!1}{2\rho}\!\right)\!\!,
\end{equation}
A substitution of \eqref{Eq:gm} into \eqref{Eq:WL_SINR_PDF_Nrx2_4}  completes the proof of \textit{Corollary 1.1}.  $\hfill\blacksquare$

\textit{Remark 5:} The analytic PDF for the $N_r \times 2$ MIMO system, that is, ${f_{{N_r} \!\times\! 2}}(\tau_j)$ in \eqref{Eq:WL_SINR_PDF_Nrx2_4}, is represented {\color{black}{as an}} infinite series, whose absolute convergence is proved by the ratio test provided in Appendix \ref{App:AppendixC}. As indicated by \eqref{eq:convspeed}, the convergence rate of the infinite series depends on the value of the SINR $\tau_j$. A faster convergence can be achieved when $\tau_j$ becomes smaller, so that the analytic PDF ${f_{{N_r} \!\times\! 2}}(\tau_j)$ is easier to calculate in the low SINR region.

\textit{Corollary 1.2:} Given the WLMMSE receiver in \eqref{Eq:WL_EST_EXPR3a}, the analytic PDF of the SINR $\tau_j$ for the $N_r \times 3 $ MIMO system, denoted by ${f_{{N_r} \!\times\! 3}}(\tau_j)$, can be expressed by \eqref{Eq:WL_SINR_PDF_Nrx3} at the bottom of the page.

\textit{Remark 6:} The detailed proof of \textit{Corollary 1.2} is similar to the analysis from \eqref{Eq:lambdaPDF} to \eqref{Eq:gm}, which is omitted here for simplicity. By comparing \eqref{Eq:WL_SINR_PDF_Nrx2_5} and \eqref{Eq:WL_SINR_PDF_Nrx3}, we find that although both ${f_{{N_r} \!\times\! 2}}(\tau_j) $ and ${f_{{N_r} \!\times\! 3}}(\tau_j) $ share a same confluent hypergeometric function $U \left(\cdot,\cdot ,\frac{\tau_j+1}{2\rho} \right) $, the form of ${f_{{N_r} \!\times\! 3}}(\tau_j)$ is more complicated. {\color{black}This is because, during the derivation of ${f_{{N_r} \!\times\! 3}}(\tau_j)$, the multivariate confluent hypergeometric function $\Phi _2^{(n)}(\cdot;\cdot;\cdot)$ is expanded as a two-fold infinite sum, and an additional infinite sum appears in the joint PDF of $f_{\bf \Lambda}({\lambda _1},{\lambda _2},\ldots,{\lambda_{N_t-1}})$ due to the Taylor series expansion on an extra exponential term ${e^{-\lambda_k}}$. They together lead to the four-fold infinite sum in the expression of ${f_{{N_r} \!\times\! 3}}(\tau_j)$, as shown in \eqref{Eq:WL_SINR_PDF_Nrx3}. } 

\section{Approximate SINR Distribution Analysis and Its Applications}
In general, the integral of a combination of confluent hypergeometric functions is mathematically intractable, which limits the applicability of the analytic PDFs given in \textit{Corollary 1.1} and \textit{Corollary 1.2}, that is, ${f_{{N_r} \!\times\! 2}}(\tau_j)$ and  ${f_{{N_r} \!\times\! 3}}(\tau_j)$, in practical MIMO performance metrics, such as outage probability and SER. Besides, the approach used to obtain ${f_{{N_r} \!\times\! 2}}(\tau_j)$ and  ${f_{{N_r} \!\times\! 3}}(\tau_j)$, unfortunately, does not fit the more general occasion, i.e., $N_t>3$. In this regard, we next exploit the moment generating function (MGF) of a gamma RV to derive an approximate PDF of the SINR $\tau_j$ for an arbitrary size WLMMSE MIMO system, summarized by the following theorem.

	\textit{Theorem 2:} Given the WLMMSE receiver in \eqref{Eq:WL_EST_EXPR3a}, an approximate analytic PDF of the SINR $\tau_j$, denoted by $f_a(\tau_j)$, under the condition that either the number of receive antennas $N_r$ is sufficiently larger than ${(N_t-1)}/{2}$ or the SNR $\rho$ is sufficiently high, can be expressed in a concise and closed form as
	\begin{align}
	\setcounter{equation}{24}
	f_a(\tau_j) &\approx  \left\{1 -{\frac{{{N_t} - 1}}{{2{N_r} - {N_t}}}\left( {\frac{{2{N_r} - {N_t} - 1}}{{2{\tau _j}}} - \frac{1}{{2\rho }}} \right)} \right\}  \nonumber \\ &\times \frac{{{{({\tau_j})}^{\frac{{2{N_r} - {N_t} - 1}}{2}}}{e^{ - \frac{{{\tau_j}}}{{2\rho }}}}}}{{{{(2\rho )}^{\frac{{2{N_r} - {N_t} + 1}}{2}}}\Gamma \left( {\frac{{2{N_r} - {N_t} + 1}}{2}} \right)}}.
	\label{Eq:finalapproxPDF}
	\end{align}

\textit{Proof:}
As stated in \textit{Remark 1}, the SINR ${\tau_j}$ is composed by summing up $N_t$ gamma RVs, therefore, by definition \cite{Moschopoulos1985}, its conditional MGF, given the eigenvalues $\lambda_1,\lambda_2,...,\lambda_{N_t-1}$, can be expressed as
\begin{align}
\!\!G(s&|{\lambda _1},{\lambda _2},\ldots,{\lambda_{N_t-1}}) \!\triangleq\! \prod\limits_{k = 1}^{N_t} {({1 -{\beta _k}s} )}^{-{\alpha _k}},
\label{Eq:WL_MGF_EXPR1}
\end{align}
when the complex frequency of the Laplace transform $s<{\rm{min}}\{ 1/\beta_1, ..., 1/\beta_{N_t}\}$.  After interpreting the shape parameters, ${\alpha _1},\ldots,{\alpha _{N_t}}$, and the scale parameter ${\beta_{N_t}} $ via \eqref{Eq:WL_SINR_PARA_EXPR1a} and \eqref{Eq:WL_SINR_PARA_EXPR1}, we have
\begin{align}
\!\!G(s&|{\lambda _1},{\lambda _2},\ldots,{\lambda_{N_t-1}}) \nonumber \\
&\!=\!  (1-{2\rho}s)^{-\frac{2N_r-N_t+1}{2}}\prod\limits_{k=1}^{N_t-1}\left({1-{\beta _k}s}\right)^{-\frac{1}{2}}\!.
\label{Eq:WL_MGF_EXPR15}
\end{align}
%

To make \eqref{Eq:WL_MGF_EXPR15} more mathematically tractable, we apply the first-order Taylor series expansion on the $k$th factorial within the sequence product $\prod\limits_{k=1}^{N_t-1}\left({1-{\beta _k}s}\right)^{-\frac{1}{2}}$, to give
\begin{equation}\label{Eq:Taloyexp}
\left({1-{\beta _k}s}\right)^{-{\frac{1}{2}}} \!=\! 1+{\frac{1}{2}}{\beta_k}{s} + O(s),
\end{equation}
where $O(s)$ denotes the second and higher-power terms of $s$. Observe that the RHS of \eqref{Eq:Taloyexp} is convergent when $s<1/\beta_k$. 
Now, upon taking \eqref{Eq:Taloyexp} into \eqref{Eq:WL_MGF_EXPR15} and omitting the higher-power terms of $s$, the conditional MGF $G(s|{\lambda _1},{\lambda _2},\ldots,{\lambda_{N_t-1}})$ can be approximated as
\begin{align}
G(s&|{\lambda _1},{\lambda _2},\ldots,{\lambda_{N_t-1}}) \nonumber \\
& \approx  (1-2\rho s)^{-\frac{2N_r-N_t+1}{2}}\left( {1 + \frac{s}{2}\sum_{k=1}^{N_t-1}\beta_k} \right).
\label{Eq:WL_MGF_EXPR2}
\end{align}
Based on \eqref{Eq:WL_MGF_EXPR2} and \eqref{Eq:REAL_WISHART_DIST}, the approximate MGF of ${\tau_j}$ can be derived by marginalization as

\begin{align}
\!\!\!G(s) &= \int\limits_0^{\infty}   \int\limits_{0}^{\lambda_1} \cdots \!\!\!\int\limits_{0}^{\lambda_{N_t-2}}  G(s|{\lambda _1},{\lambda _2},\ldots,{\lambda_{N_t-1}}) \nonumber \\
&\times f_{\bf \Lambda}({\lambda _1},{\lambda _2},\ldots,{\lambda_{N_t-1}})d\lambda_{N_t-1}\cdots d\lambda_2d\lambda_1  \nonumber  \\
&= (1-2\rho s)^{-\frac{2N_r-N_t+1}{2}} \left( {1 + G_0s} \right),
\label{Eq:WL_MGF_EXPR4}
\end{align}
where
\begin{align}
G_0 &= \int\limits_0^{\infty}   \int\limits_{0}^{\lambda_1} \cdots \!\!\!  \int\limits_{0}^{\lambda_{N_t-2}} \; \frac{1}{2}\sum_{k=1}^{N_t-1}\beta_k f_{\bf \Lambda}({\lambda _1},{\lambda _2},\ldots,{\lambda_{N_t-1}})  \nonumber \\
&\times d\lambda_{N_t-1}\cdots d\lambda_2d\lambda_1.	
\label{Eq:WL_MGF_EXPR5}
\end{align}
Moreover, based on \eqref{Eq:WL_SINR_PARA_EXPR1}, the scale parameter $\beta_k$ in \eqref{Eq:WL_MGF_EXPR5} can be further approximated as
\begin{equation}\label{Eq:approximation}
\beta_k =  (\lambda_k+\frac{1}{2}\rho^{-1})^{-1}\approx \lambda_k^{-1},
\end{equation}
when the eigenvalue $\lambda_k \gg \frac{1}{2}\rho^{-1}$. This is a mild condition in practice, easily achievable either when the SNR $\rho$ is sufficiently high or when the number of receive antennas $N_r$ is sufficiently greater than\footnote{For a real-valued Wishart matrix $  {\mathcal{W}}_{N_t-1}(\frac{1}{2}{\bf I}_{N_t-1},2N_r)$, the expectation of its minimum eigenvalue is proven to monotonically increase with $2N_r-N_t+1$ \cite{Edelman1991}.} ${(N_t-1)}/{2}$. With the approximated $\beta_k$ in hand, the quantity $G_0$ in \eqref{Eq:WL_MGF_EXPR5} becomes
\begin{align}
G_0 &\approx \frac{1}{2}\int\limits_0^{\infty}   \int\limits_{0}^{\lambda_1} \cdots \!\!\!\int\limits_{0}^{\lambda_{N_t-2}}  \left(\sum_{k=1}^{N_t-1}\frac{1}{\lambda_k}\right)
f_{\bf \Lambda}({\lambda _1},{\lambda _2},\ldots,{\lambda_{N_t-1}}) \nonumber \\
&d\lambda_{N_t-1} \cdots d\lambda_2d\lambda_1\nonumber\\
&= 	\frac{1}{2}E\big[{\rm Tr}\{{\widetilde{\bf C}}^{-1}\}\big]=\frac{1}{2}{\rm Tr}\big\{E[{\widetilde{\bf C}}^{-1}]\big\},\label{Eq:WL_MGF_G0approx1}
\end{align}
in which $ {\widetilde{\bf C}} \sim {\mathcal{W}}_{N_t-1}(\frac{1}{2}{\bf I}_{N_t-1},2N_r) $ is a real-valued Wishart matrix{\color{black}, with $\lambda_{1},\lambda_{2},\ldots,\lambda_{N_t-1}$ being its eigenvalues.} According to \cite[Theorem 3.1 (i)]{Rosen1988}, the expectation of $ {\widetilde{\bf C}^{-1}}$ can be obtained as
\begin{equation}\label{eq:theoRosen1988}
E\big[{\widetilde{\bf C}}^{-1}\big]=\frac{2}{2N_r-N_t}{\bf I}_{N_t-1}.
\end{equation}
Upon inserting \eqref{eq:theoRosen1988} into \eqref{Eq:WL_MGF_G0approx1}, the quantity $G_0$ can be finally simplified as
\begin{equation}
G_0\approx\frac{1}{2}{\rm Tr}\left\{\frac{2}{2N_r-N_t}{\bf I}_{N_t-1}\right\}=\frac{N_t-1}{2N_r-N_t}.
\label{Eq:WL_MGF_G0approx}
\end{equation}
A substitution of \eqref{Eq:WL_MGF_G0approx} into \eqref{Eq:WL_MGF_EXPR4} gives
\begin{align}\label{Eq:finalMGF}
G(s) \!\approx\!
(1\!-\!2\rho s)^{-\frac{2N_r-N_t+1}{2}} \!\!-\!\! {\frac{(N_t\!-\!1)(-s)}{2N_r\!-\!N_t}}(1\!-\!2\rho s)^{-\frac{2N_r-N_t+1}{2}}.
\end{align}
After applying the inverse Laplace transform on $G(s)$ given in Appendix \ref{App:invLaplace}, we complete the proof of \textit{Theorem 2}. $\hfill\blacksquare$

\textit{Remark 7:} By comparing the approximate PDF, $f_a(\tau_j)$, given in \textit{Theorem 2} with \eqref{Eq:lemma1} in Appendix \ref{App:GammaRV}, we observe that $f_a(\tau_j)$ is the product of the term, $1 -{\frac{{{N_t} - 1}}{{2{N_r} - {N_t}}}\left( {\frac{{2{N_r} - {N_t} - 1}}{{2{\tau _j}}} - \frac{1}{{2\rho }}} \right)} $, and the PDF of a gamma RV with a shape parameter, $ \frac{{2{N_r} - {N_t} - 1}}{2}$, and a scale parameter, $2\rho$. Since a gamma RV gradually turns to be Gaussian when its shape parameter approaches to infinity \cite{leemis2008univariate}, it is expected that $f_a(\tau_j)$ becomes a more Gaussian-like PDF as $2N_r-N_t$ increases. Besides, the derived approximate PDF $f_a(\tau_j)$ is also valid for massive antenna systems, as  large $N_r$ and $N_t$ further relax the mild condition {\color{black}{for}} the approximation in \eqref{Eq:approximation}.

We next demonstrate the usefulness of \textit{Theorem 2} in some practical MIMO system performance metrics, including outage probability, SER, and diversity gain, which theoretically explains the simulated and experimental results in previous studies on WLMMSE estimators \cite{Kuchi2009,Deng2019,YangLamare2015}. The results are summarized by the following three corollaries.


\textit{Corollary 2.1:}  Given a target rate $R$, the outage probability of the $j$th spatial stream of the WLMMSE receiver in \eqref{Eq:WL_EST_EXPR3a}, denoted by  ${\rm{P}}_{{\rm{out}},j}(R)$, can be approximated as
\begin{align}
&{\rm{P}}_{{\rm{out}},j}(R) \!  \approx \! \frac{\gamma\left(\!\frac{{2{N_r} - {N_t} + 1}}{2},\!\frac{ {2^R-1} } {2\rho}\right)\! }{\Gamma \left(\frac{2N_r - N_t + 1}{2}\right)} \nonumber \\
& -  \frac{N_t-1}{2N_r-N_t}\frac{e^{- \frac{{2^R-1}}{2\rho}}}{ {2^R-1}}\frac{({\frac{{2^R-1}}{2\rho}})^{\frac{2N_r-N_t+1}{2}} } { \Gamma \left( {\frac{{2{N_r} - {N_t} + 1}}{2}} \right)}.
\label{Eq:WL_CDF_EXPR1}
\end{align}
\textit{Proof:}
The outage probability of a data stream is defined as the probability that its output SINR cannot support the assigned target rate \cite{simon2005digital}. {\color{black}Suppose all the spatial streams adopt independent encoding schemes with the same rate $R$, we have}
\begin{align}
&{\rm{P}}_{{\rm{out}},j}(R) \triangleq {\rm{Pr}}\big({\rm log}_2(1+\tau_j)<R\big),
\label{Eq:OT_DEFINATION}
\end{align}
which can be computed by integrating the approximate PDF  $f_a(\tau_j)$, to give 
\begin{align}
& {\rm{Pr}}\big({\rm log}_2(1+\tau_j)<R\big) \approx \int\limits_0^ { 2^R-1 } f_a(\tau_j) d\tau_j.
\label{Eq:OT_F_FUNC}
\end{align}
%
Upon inserting  \eqref{Eq:finalapproxPDF} into \eqref{Eq:OT_F_FUNC}, and after some algebraic manipulations, we arrive at \eqref{Eq:WL_CDF_EXPR1}. This completes the proof of \textit{Corollary 2.1}. $\hfill\blacksquare$

\textit{Remark 8:}
Recall that the outage probability of the $j$th spatial stream of the widely linear zero-forcing (WLZF) detector, denoted by ${\rm{P}}_{{\rm{out}},j}^{\rm{WLZF}}(R)$, obeys the form \cite{xu2006performance,Ducoing2016}
\begin{align}
&{\rm{P}}_{{\rm{out}},j}^{\rm{WLZF}}(R)= \frac{\gamma\left(\frac{{2{N_r} - {N_t} + 1}}{2},\frac{ 2^R-1 } {2\rho}\right) }{\Gamma \left(\frac{2N_r - N_t + 1}{2}\right)}.
\label{Eq:WL_WLZF_CDF_EXPR2}
\end{align}
{\color{black}{This}} normalized incomplete gamma function {\color{black}{is}} identical to the first term on the RHS of \eqref{Eq:WL_CDF_EXPR1}, and it indicates that the outage probability of the WLMMSE detector, ${\rm{P}}_{{\rm{out}},j}(R)$ in \eqref{Eq:WL_CDF_EXPR1}, consists of two parts, the outage probability of the WLZF detector (the first term), ${\rm{P}}_{{\rm{out}},j}^{\rm{WLZF}}(R)$, and the difference between them (the second term). In addition, we can observe that the difference term approaches to zero as the target rate $R$ increases. This finding is in analogy to the relationship between the linear ZF and LMMSE detectors {\color{black}{discussed}} in \cite{Lim2019}.

\textit{Corollary 2.2:}  When the transmit rectilinear or QR signal conveys one bit per symbol, such as BPSK or OQPSK modulated, the average symbol error rate on the $j$th spatial stream of the WLMMSE receiver in \eqref{Eq:WL_EST_EXPR3a} {\color{black}{is}}

%
\begin{align}
P_e(\rho) &\approx  {\frac{4N_r - 3N_t + 1}{12(4N_r-2N_t)}} {\left( {1 + \rho } \right)^{ - \frac{{2{N_r} - {N_t} + 1}}{2}}} \nonumber \\
& +  {\frac{{6{N_r} - 5{N_t} + 2}}{4(6N_r - 3N_t)}} {\left( {1 + \frac{4}{3}\rho } \right)^{ - \frac{{2{N_r} - {N_t} + 1}}{2}} }.
\label{Eq:WL_BER_FUNC_APPR2}
\end{align}

\textit{Proof:}
The SER represents the probability of transmitting a correct symbol but erroneously receiving its distorted version, which, based on the law of total probability, can be calculated by marginalization as
\begin{gather}
P_e(\rho) = \int\limits_0^\infty  P_e(\rho | \tau_j ) f(\tau_j) d{\tau_j},\label{Eq: WL_BER_FUNC_APPR3}
\end{gather}
where $P_e(\rho|\tau_j)$ is the conditional SER given the SINR $\tau_j$. Based on the WLMMSE estimator in \eqref{Eq:WL_EST_EXPR3a}, the conditional SER  $P_e(\rho|\tau_j)$ for OQPSK signals can be obtained as \cite[Eq. (5-2-5)]{Proakis2001}
\begin{gather}
P_e(\rho|\tau_j)  = {\mathcal Q}\left( {\sqrt {\frac{{E\left[ {{{\left|{\epsilon_j}{x_j} \right|}^2}} \right]}}{{E\left[ {{{\left| {\xi _j}  \right|}^2}} \right]}}} } \right) = {\mathcal Q}\left( {\sqrt {\tau_j}   } \right).
\label{Eq:WL_BER_EXPR_1}
\end{gather}
By considering a numerical approximation of the Q-function, that is \cite{Chiani2003}
\begin{gather}
{\mathcal Q}\left( x \right) \approx \frac{1}{{12}}{e^{ - \frac{{{x^2}}}{2}}} + \frac{1}{4}{e^{ - \frac{{2{x^2}}}{3}}},
\label{Eq: WL_Q_FUNC_APPR1}
\end{gather}
the conditional SER $P_e(\rho|\tau_j)$ in \eqref{Eq:WL_BER_EXPR_1} becomes
\begin{gather}
P_e(\rho|\tau_j) \approx \frac{1}{{12}}{e^{ - \frac{1}{2}\tau_j   } } + \frac{1}{4}{e^{ - \frac{{2}}{3}\tau_j}}.
\label{Eq:BER_FUNC_APPR1}
\end{gather}
Upon substituting \eqref{Eq:BER_FUNC_APPR1} into \eqref{Eq: WL_BER_FUNC_APPR3}, we have

\begin{align}\label{Eq:SERderivation}
P_e(\rho) &= \int\limits_0^\infty  P_e(\rho|\tau_j)   \int\limits_0^{\infty}   \int\limits_{0}^{\lambda_1} \cdots \!\!\!\int\limits_{0}^{\lambda_{N_t-2}}  f({\tau_j}|{\lambda _1},{\lambda _2},...,{\lambda _{{N_t} - 1}}) \nonumber \\
&\times  f_{\bf \Lambda}({\lambda _1},\lambda_2,...,{\lambda_{N_t-1}}) d\lambda_{N_t-1}\cdots d\lambda_2d\lambda_1 \nonumber \\
&\approx \int\limits_0^{\infty}    \int\limits_{0}^{\lambda_1}  \cdots   \int\limits_{0}^{\lambda_{N_t-2}}
\left( \int\limits_0^\infty {(\frac{1}{12}{e^{ - \frac{1}{2}\tau_j} }  +  \frac{1}{4}{e^{ - \frac{{2}}{3}\tau_j}} )} \right. \nonumber \\
&\left. \times \;\;  f({\tau_j}| {\lambda _1},{\lambda _2},...,{\lambda _{{N_t} - 1}} ) d{\tau_j}  \right)
\nonumber \\
&\times \;\; f_{\bf \Lambda}({\lambda _1},{\lambda _2},...,{\lambda_{N_t-1}}) d\lambda_{N_t-1}\cdots d\lambda_2d\lambda_1  \nonumber  \\
&= \int\limits_0^{\infty}  \int\limits_{0}^{\lambda_1} \cdots   \int\limits_{0}^{\lambda_{N_t-2}} \left[ \frac{1}{{12}}{G}\left(  { - \frac{1}{2}|{\lambda _1},{\lambda _2},...,{\lambda _{{N_t} - 1}} }  \right)  \right. \nonumber \\
&\left. + \frac{1}{4}{G}\left(  { - \frac{2}{3}|{\lambda _1},{\lambda _2},...,{\lambda _{{N_t} - 1}} }  \right) \right]
\nonumber \\
&\times f_{\bf \Lambda}({\lambda _1},{\lambda _2},...,{\lambda_{N_t-1}}) d\lambda_{N_t-1}\cdots d\lambda_2d \lambda_1 .
\end{align}
Now, upon substituting \eqref{Eq:finalMGF} and \eqref{Eq:REAL_WISHART_DIST} into \eqref{Eq:SERderivation}, and after some manipulations, we complete the proof of \textit{Corollary 2.2}. $\hfill\blacksquare$


\textit{Corollary 2.3:} The diversity gain of the considered MIMO system with the WLMMSE receiver in \eqref{Eq:WL_EST_EXPR3a}, denoted by $\mathcal{D}$, {\color{black}{is}}
\begin{equation}\label{Eq:WL_DIVERSITY_EQ4}
\mathcal{D}=N_r-\frac{N_t-1}{2}.
\end{equation}

\textit{Proof:} The diversity gain $\mathcal{D}$ of a general MIMO system is defined as \cite{Tse2003}
\begin{gather}
\mathcal{D} = - \lim_{\rho  \to \infty } \frac{ \log P_e(\rho) }{{\log \rho }}.
\label{Eq:WL_DIVERSITY_EQ1}
\end{gather}

According to \cite[Eq. (5.3)]{simon2005digital}, the SER  $P_e(\rho)$ can be expressed in terms of the MGF of $\tau_j$, $G(s)$, as
\begin{equation}
P_e(\rho) = \frac{1}{\pi }\int\limits_0^{\pi /2} {{{G}}\left( - \frac{1}{2{{{\sin }^2}\theta }}\right)} d\theta.
\label{Eq:SER_EXPR_2}
\end{equation}

Upon substituting \eqref{Eq:SER_EXPR_2} into \eqref{Eq:WL_DIVERSITY_EQ1} and exploiting \eqref{Eq:WL_MGF_EXPR4}, the diversity gain $\mathcal{D}$ can be derived as
\begin{align}
\!&\mathcal{D} = - \lim_{\rho  \to \infty } \frac{1}{\log \rho}\times  {\log \Bigg \{ \frac{1}{\pi }\int\limits_0^{\pi /2} {{{G}}\left( - \frac{1}{2{{{\sin }^2}\theta }}\right)}  d\theta}  \Bigg \} \nonumber\\
&= \!\!-\!\!\int\limits_0^{\infty} \!\! \int\limits_{0}^{\lambda_1} \!\! \cdots\!\!\!\!\!\! \int\limits_{0}^{\lambda_{N_t-2}} \!\!\!\!\!\lim_{\rho  \to \infty }\! \underbrace{\frac{{\log \bigg \{ \frac{1}{\pi }\int\limits_0^{\pi /2}  G\left( { - \frac{1}{{2{{\sin }^2}\theta }}|{\lambda _1}, \ldots ,{\lambda _{{N_t} - 1}}} \right)d\theta \bigg \} }}{{\log \rho }}}_{L(\rho)}\nonumber \\
& \times f_{\bf \Lambda}({\lambda _1},{\lambda _2},\ldots,{\lambda_{N_t-1}})d\lambda_{N_t-1}\cdots d\lambda_2d\lambda_1.
\label{Eq:WL_DIVERSITY_EQ2}
\end{align}
By further considering \eqref{Eq:WL_MGF_EXPR15}, the term $\mathop {\lim }\limits_{\rho  \to \infty } L(\rho)$ in \eqref{Eq:WL_DIVERSITY_EQ2} becomes
\begin{align}
&\lim_{\rho \to \infty} L(\rho) = \mathop {\lim }\limits_{\rho  \to \infty } \frac{1}{{{\log \rho }}} \times \log \nonumber \\
&\left\{{\rho ^{ - \frac{{2{N_r} - {N_t} + 1}}{2}}}
\times \frac{1}{\pi } \int\limits_0^{\pi /2}  {{\left(\frac{1}{\rho } + \frac{1}{{{{\sin }^2}\theta }}\right)}^{ - \frac{{2{N_r} - {N_t} + 1}}{2}}} \right. \nonumber \\
&\left. \times \prod\limits_{k = 1}^{{N_t} - 1} {{{\left( {1 + \frac{1}{{\left( {2{\lambda _k} + {\rho ^{ - 1}}} \right){{\sin }^2}\theta }}} \right)}^{ - 1/2}}}  d\theta  \right\}
 \nonumber \hfill \nonumber \\
&=  -{N_r} + \frac{{{N_t} - 1}}{2} + \mathop {\lim }\limits_{\rho  \to \infty } \frac{1}{{\log \rho }} \times \log \nonumber \\
& \left\{ \frac{1}{\pi }\int\limits_0^{\pi /2} \!\!{{\left(\frac{1}{{{{\sin }^2}\theta }}\right)}^{ - \frac{{2{N_r} - {N_t} + 1}}{2}}}  \!\times\! \prod\limits_{k = 1}^{{N_t} - 1} \!\!{{{\left( {1 + \frac{1}{{{2{\lambda _k}} {{\sin }^2}\theta }}} \right)}^{ - 1/2}}}  d\theta  \right\} \nonumber \\
&= -N_r+\frac{N_t-1}{2}.
\label{Eq:WL_DIVERSITY_EQ3}
\end{align}
A substitution of \eqref{Eq:WL_DIVERSITY_EQ3} into \eqref{Eq:WL_DIVERSITY_EQ2} completes the proof of \textit{Corollary 2.3}. $\hfill\blacksquare$
%
%

\textit{Remark 9:} Unlike the outage probability ${\rm{P}}_{{\rm{out}},j}(R)$ in \eqref{Eq:WL_CDF_EXPR1} and the SER $P_e(\rho)$ in \eqref{Eq:WL_BER_FUNC_APPR2}, the diversity gain $\mathcal{D}$ in \eqref{Eq:WL_DIVERSITY_EQ4} is an exact value for WLMMSE MIMO systems since its derivation does not involve any approximation. It is also the decaying rate of the exponential functions involved in the SER $P_e(\rho)$ in \eqref{Eq:WL_BER_FUNC_APPR2}. Recall that the diversity gain for the conventional LMMSE MIMO system is given by $\mathcal{D}_{\scriptscriptstyle {\text {LMMSE}}}=N_r-N_t+1$ \cite{Kim2008,Lim2019}, hence the diversity gain improvement introduced by a WLMMSE detector, denoted by $\Delta \mathcal{D}$, can be quantified as
\begin{equation}\label{Eq:Deltadiversitygain}
\Delta \mathcal{D} \triangleq \mathcal{D}-\mathcal{D}_{\scriptscriptstyle {\text {LMMSE}}} = \frac{N_t-1}{2},
\end{equation}
and this improvement linearly increases with the number of transmit antennas $N_t$.
%
\begin{table}
	
	{\color{black}
		\centering
		\caption{Approximations Used in Section IV}
		\label{tab:approx}
		
		\scalebox{0.88}
		{	
			\begin{tabular}{|c|c|c|c|}
				\hline
				Eq. & Expression & Approximation & Method / Condition   \\ \hline
				\eqref{Eq:WL_MGF_EXPR2}  & $\prod\limits_{k=1}^{N_t-1}\left({1-{\beta _k}s}\right)^{-\frac{1}{2}}$  &  ${1 + \frac{s}{2}\sum_{k=1}^{N_t-1}\beta_k}$ & Taylor series expansion \\ \hline
				\eqref{Eq:approximation} & $\lambda_k+\frac{1}{2}\rho^{-1}$  & $\lambda_k$ &  $\rho \gg 0 $ or   $N_r \gg \frac{N_t-1}{2} $   \\ \hline
				\eqref{Eq: WL_Q_FUNC_APPR1} & ${\mathcal Q}\left( x \right)$  & $\frac{1}{{12}}{e^{ - \frac{{{x^2}}}{2}}} + \frac{1}{4}{e^{ - \frac{{2{x^2}}}{3}}}$ & Numerical approximation   \\ \hline
			\end{tabular}
		}
	}
\end{table}
\begin{figure}
	\centering
	\includegraphics[width=0.95\linewidth]{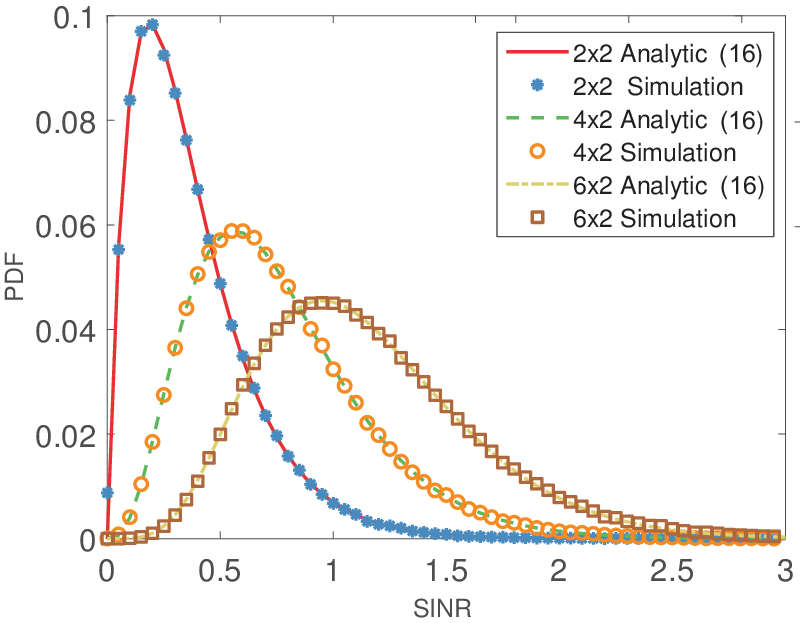}
	\caption{Analytic and simulated PDFs ${f_{{N_r} \!\times\! 2}}(\tau_j)$ as functions of the SINR.}
	\label{Fig:WL_SINR_PDF_PLOT1}
\end{figure}

{\color{black}For clarity, the approximations used in the above derivations are summarized in Table \ref{tab:approx}.}

\section{Simulations}\label{Sec:sim}
Monte Carlo simulations were provided to validate our theoretical performance analysis on the considered WLMMSE MIMO systems. Throughout the simulations, the transmitted data streams ${\bf{d}}(l)$ were OQPSK-modulated. This QR constellation has been widely applied in the state-of-the-art low power wide area wireless technologies such as ZigBee, Wireless Smart Utility Network and RFID, etc \cite{Ayoub2019,Dawoud2020,Pillai2007}. The transmit power was set to $E_s = 1$, and the channel was flat Rayleigh fading.
%
\begin{figure}[htbp]
\centering
\subfloat[]{ \label{Fig:WL_SINR_PDF_MGF_PLOT2:SNR}
	\centering \includegraphics[width=\linewidth]{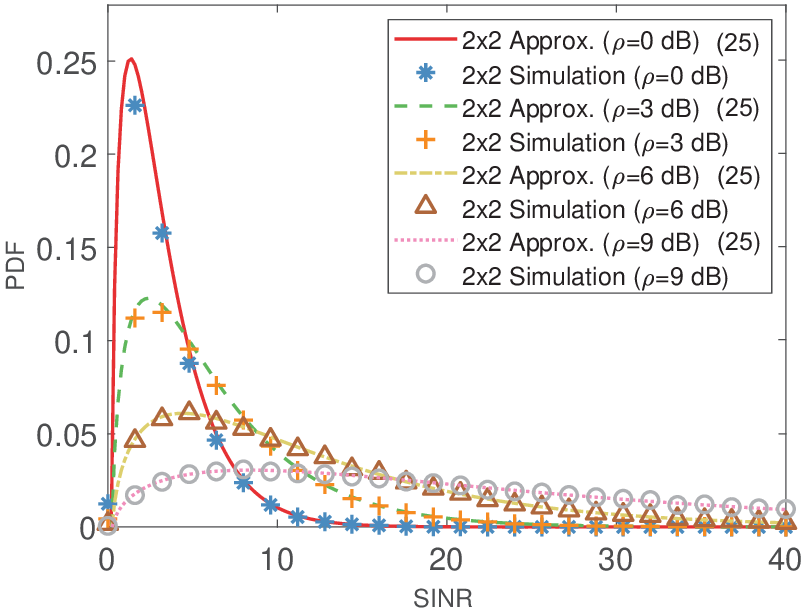}
}
\\
\subfloat[]{ \label{Fig:WL_SINR_PDF_MGF_PLOT2:SmallAntenna}
\begin{minipage}[b]{0.53\textwidth}%
	\centering \includegraphics[width=\linewidth]{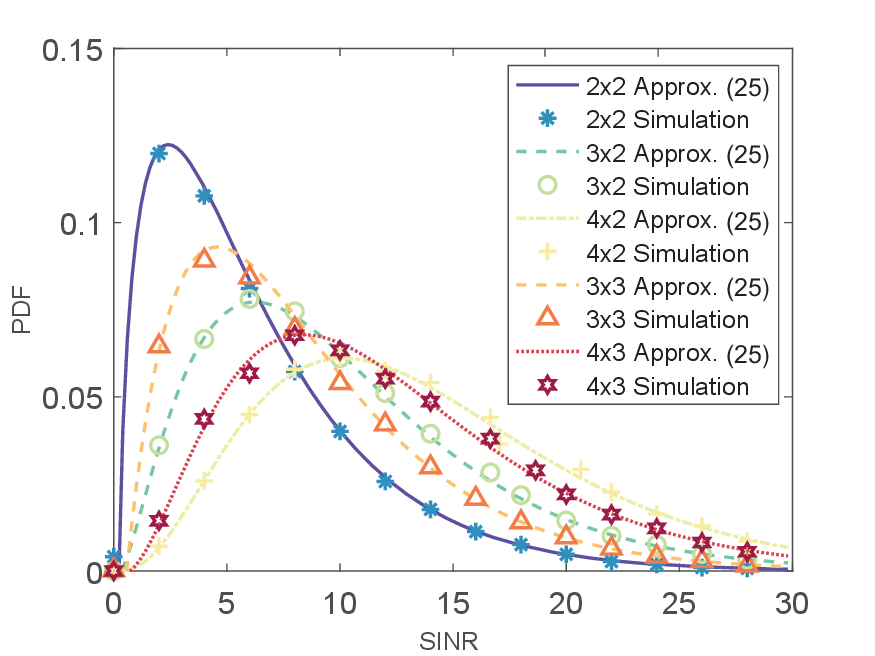} %
\end{minipage}
}
\hfill
\subfloat[]{ \label{Fig:WL_SINR_PDF_MGF_PLOT2:LargeAntenna}
\begin{minipage}[b]{0.49\textwidth}%
	\centering \includegraphics[width=\linewidth]{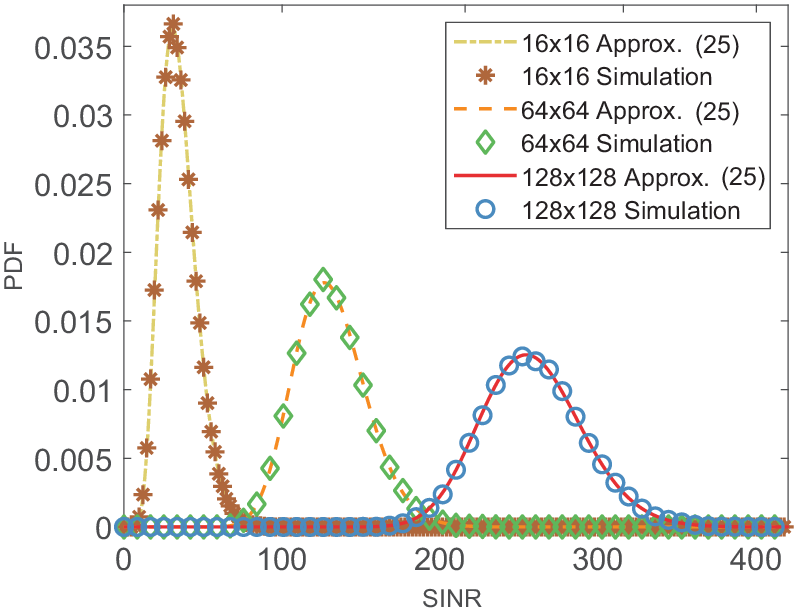} %
\end{minipage}
}
\hfill
\caption{Approximate PDF $f_a(\tau_j)$  and its simulated counterpart as functions of the SINR, against different antenna configurations and SNRs. (a) a $2\times2$ MIMO system with $\rho=$ 0 dB, 3 dB, 6 dB, 9 dB. (b) $2\times2$, $3\times2$, $4\times2$, $3\times3$ and $4\times3$ MIMO systems with $\rho=$3 dB. (c) $16\times16$, $64\times64$ and $128\times128$ MIMO systems with $\rho=$3 dB.}
\label{Fig:WL_SINR_PDF_MGF_PLOT2}
\end{figure}

In the first set of simulations, the validity of the analytic PDF of the SINR was examined for $N_r \times 2 $ MIMO systems. The SNR $\rho$ was set to -10 dB, and three different antenna configurations, $2 \times 2$, $4 \times 2$ and $6 \times 2$, were considered for illustration purpose.  The analytic ${f_{{N_r} \!\times\! 2}}(\tau_j)$ was evaluated based on \eqref{Eq:WL_SINR_PDF_Nrx2_5} in \textit{Corollary 1.1}, in which the number of convergent series used to plot ${f_{2 \!\times\! 2}}(\tau_j)$, ${f_{4 \!\times\! 2}}(\tau_j)$ and ${f_{6 \!\times\! 2}}(\tau_j)$, was 5, 15 and 25, respectively, while the simulated PDF was illustrated by a histogram calculated from $1\times10^6$ independent realizations for each antenna configuration. Observe in Fig. \ref{Fig:WL_SINR_PDF_PLOT1} that the analytic PDF spread out away from the ordinate and became broader and shallower as the number of receive antennas, $N_r$, increased from 2 to 6. This is expected, because, given a fixed number of transmit antennas $N_t$, the expectation of the SINR $\tau_j$ with respect to the channel entry $\widetilde h_{k,j}$ can be obtained from \eqref{Eq:WL_SINR_EXPR3} as
\begin{align}
E[{\tau _j}] = \rho  \left( {2{N_r} - {N_t} + 1 + \sum\limits_{k = 1}^{{N_t} - 1} {\frac{1}{{1 + 2\rho {\lambda _k}}}} } \right),
\end{align}
which monotonically increases against $N_r$.  Moreover, the analytic and simulated PDFs were well matched for different SINR situations and antenna configurations.

\begin{figure}[t]
	\centering
	\includegraphics[width=0.95\linewidth]{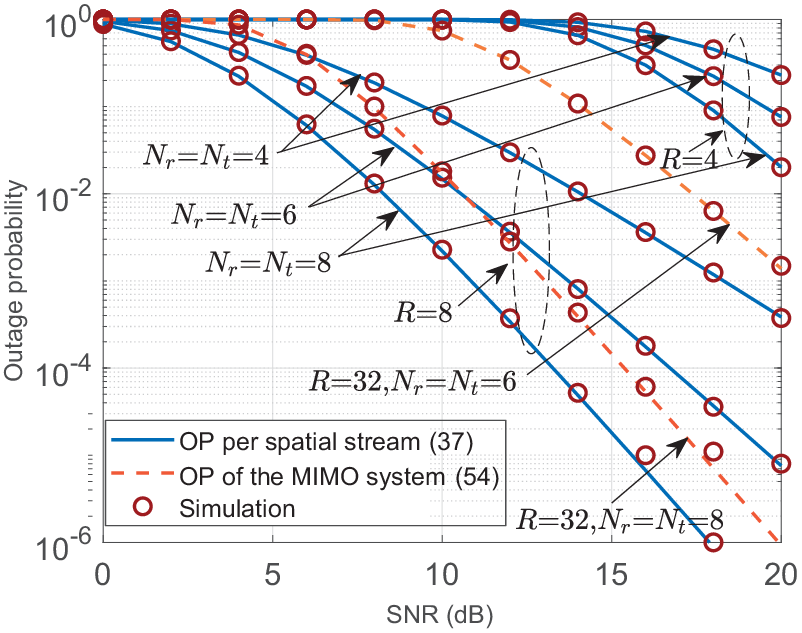}
	\caption{Analytic and simulated outage probability for different antenna configurations with $R=4$ bits/s/Hz and $R=8$ bits/s/Hz for spatial streams and $R=32$ bits/s/Hz for the whole MIMO system.}
	\label{Fig:WL_OP_PLOT1}
\end{figure}
\begin{figure}[htbp]
	\centering
	\subfloat[]{ \label{Fig:WL_Outage_PLOT4:RTXVAR}
		\centering \includegraphics[width=\linewidth]{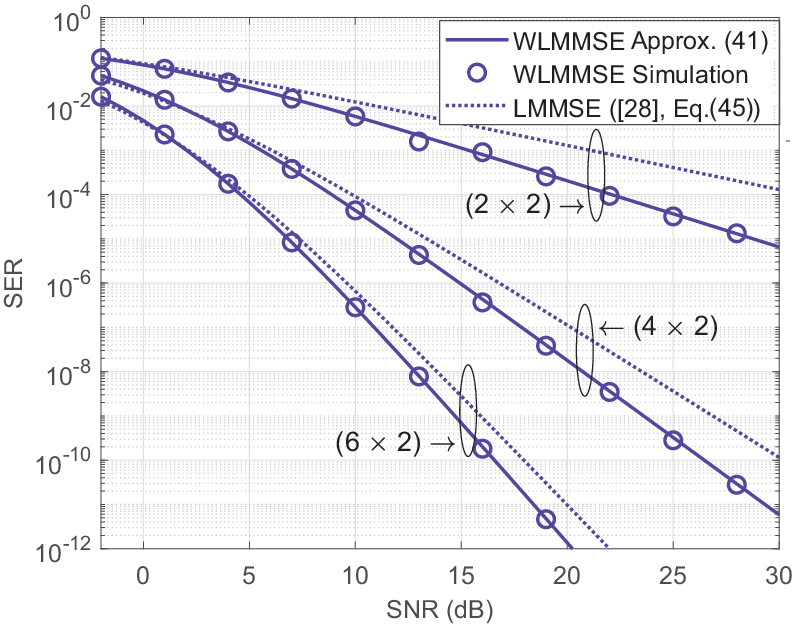}
	}
	\\
	\subfloat[]{ \label{Fig:WL_Outage_PLOT4:RXEQTX}
		\begin{minipage}[b]{0.49\textwidth}%
			\centering \includegraphics[width=\linewidth]{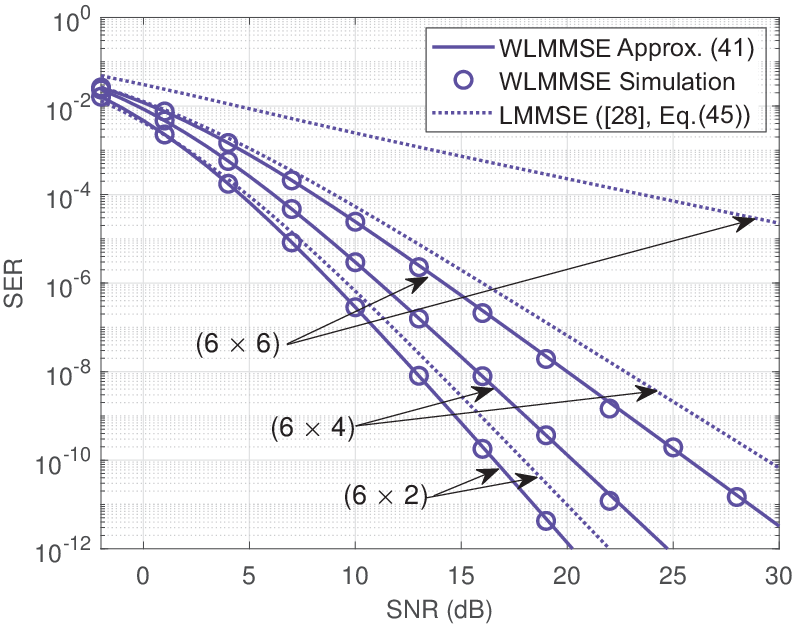} %
		\end{minipage}
	}
	\\
	\subfloat[] { \label{Fig:RXEQTX:Massive}
		\begin{minipage}[b]{0.49\textwidth}%
			\centering \includegraphics[width=\linewidth]{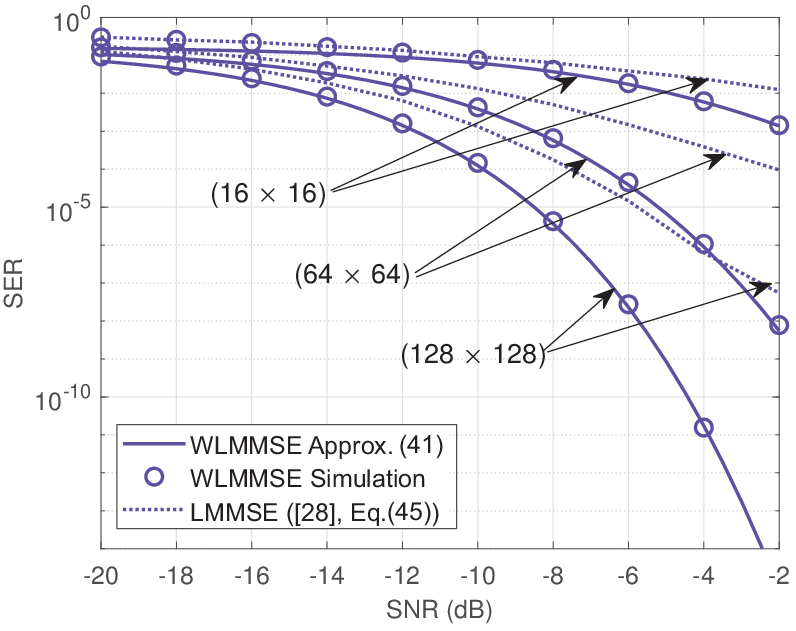} %
		\end{minipage}
	}
\caption{SERs of WLMMSE and LMMSE {\color{black}{receivers}} as functions of the SNR $\rho$. (a) $2\times2$, $4\times2$ and $6\times2$ MIMO systems. (b) $6\times2$, $6\times4$ and $6\times6$ MIMO systems. (c) $16\times16$, $64\times64$ and $128\times128$ massive MIMO systems.}
\label{Fig:SER_PLOT4}
\end{figure}

In the next stage, we compared the approximate PDF $f_a(\tau_j)$, given by \eqref{Eq:finalapproxPDF} in \textit{Theorem 2}, with its numerical counterpart, which was illustrated by a histogram computed from $8\times10^5$ independent realizations. In this experiment, we first considered a $2\times2$ MIMO system, and set the SNR $\rho=$ 0 dB, 3 dB, 6 dB, 9 dB, respectively. As shown in Fig. \subref{Fig:WL_SINR_PDF_MGF_PLOT2:SNR}, when $\rho=$ 0 dB, the mismatch between the approximate and simulated PDFs appeared around the peak of the PDF curve. However, when $\rho\ge$ 3 dB, the approximate PDFs were in good agreement with their respective numerical results in the entire SINR range. Next, we fixed the SNR $\rho=3$ dB, and examined the approximate PDF for small-scale and large-scale antenna configurations in Fig. \subref{Fig:WL_SINR_PDF_MGF_PLOT2:SmallAntenna} and Fig. \subref{Fig:WL_SINR_PDF_MGF_PLOT2:LargeAntenna}, respectively. In both the figures, the high accuracy of those approximate PDFs in predicting their empirical counterparts can be observed. Moreover, Fig. \subref{Fig:WL_SINR_PDF_MGF_PLOT2:LargeAntenna} shows that, the approximate PDF $f_a(\tau_j)$ appeared to be Gaussian-like in large-scale $64\times64$ and $128\times128$ MIMO systems, as discussed in \textit{Remark 7}. To summarize, the condition used to approximate the MGF function $G(s)$ from \eqref{Eq:WL_MGF_EXPR4} to \eqref{Eq:finalMGF}, that is, either the SNR $\rho$ is sufficiently high or the number of receive antennas $N_r$ is sufficiently larger than $\frac{N_t-1}{2}$, is rather mild, as the approximate PDF $f_a(\tau_j)$ is inaccurate only in some special cases, e.g., $\rho \le 0$ dB and $N_r=N_t=2$.
\begin{figure}[htbp]
	\centering
	\includegraphics[width=0.95\linewidth]{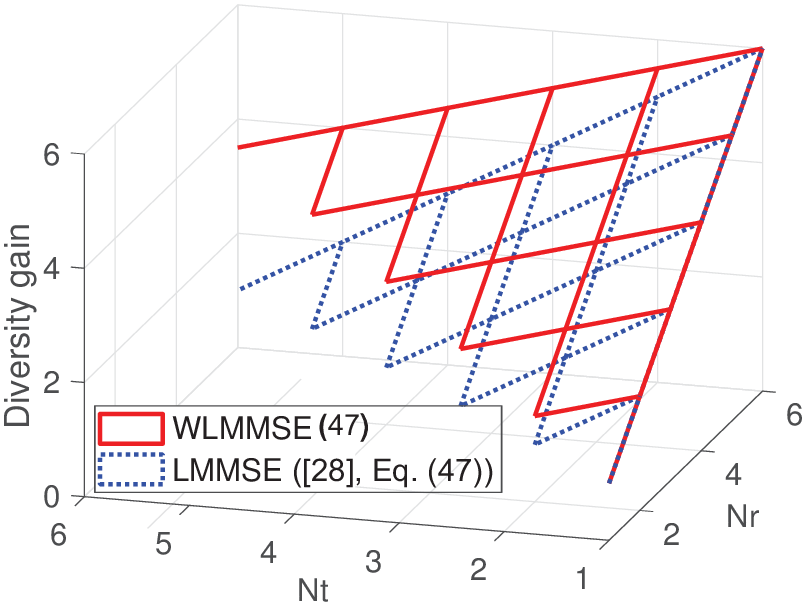}
	\caption{Diversity gains of WLMMSE and LMMSE MIMO systems with different $N_t$ and $N_r$.}
	\label{Fig:WL_Outage_PLOT4:RXEQTX:WL_DGain_PLOT}
\end{figure}

We next examined the validity of the proposed outage probability analysis in  \eqref{Eq:WL_CDF_EXPR1}. In this experiment, three different antenna configurations, $4 \times 4$, $6 \times 6$, $8 \times 8$, were considered, and in each setting, two different target rates, $R$ = 4 bits/s/Hz  and $R$ = 8 bits/s/Hz, were used. As compared, the outage probability of the whole MIMO system was also considered, given by
\begin{equation}
{{\rm{P}}_{{\rm{out}}}}(R) = 1 \! -\! {\left[ {1 - {{\rm{P}}_{{\rm{out}},j}}\left(\frac{R}{{{N_t}}}\right)} \right]^{{N_t}}}.
\end{equation}
Its numerical examples were conducted in $6 \times 6$ and $8 \times 8$ MIMO systems with a target rate $R=$ 32 bits/s/Hz. Each numerical result was obtained by averaging $1 \times 10^8$ independent Rayleigh fading channel realizations. As  shown  in  Fig. \ref{Fig:WL_OP_PLOT1}, a good agreement can be observed between each analytic outage probability curve and its empirical counterpart. Besides, as expected, an improvement in the outage probability can be obtained either by using a higher target rate $R$ or by considering a larger antenna configuration.

In the next set of simulations, the validity of the approximate SER $P_e(\rho)$ in \eqref{Eq:WL_BER_FUNC_APPR2} was investigated for five small-scale and three large-scale MIMO systems. The analytic SER of the LMMSE {\color{black}{receiver}} \cite[Eq. (45)]{Lim2019} was also illustrated for comparison. The simulation results in Fig. \ref{Fig:SER_PLOT4} were obtained by averaging $1\times10^8$ independent realizations, each with $1\times10^4$ transmitted data symbols. They can be interpreted in two aspects. First, for all antenna configurations, the approximate SER of the WLMMSE receiver was always closely matched with its empirical counterpart. Second, the WLMMSE receiver provided considerable SER gains over the LMMSE one in different situations. This is expected, because the WL processing framework possesses the modeling advantage over the strictly linear one for the considered MIMO transmission system with QR constellation mappings.

Finally, the diversity gain of the WLMMSE MIMO system, given in \textit{Corollary 2.3}, is plotted in Fig. \ref{Fig:WL_Outage_PLOT4:RXEQTX:WL_DGain_PLOT} as a function of different numbers of transmit and receive antennas. Its improvement over the LMMSE one \cite{Lim2019}, independent on the number of receive antennas $N_r$, linearly increased with the number of transmit antennas $N_t$, which echoes the analysis in \textit{Remark 9}.

\section{Conclusion and Outlook}
{\color{black}{We have introduced a}} post-detection SINR analysis of WLMMSE MIMO systems with rectilinear or QR signals over uncorrelated Rayleigh fading channels. We have derived first an analytic PDF {\color{black}{of the SINR}} for WLMMSE MIMO systems with $N_t=2,3$ and subsequently, an approximate closed form PDF valid for {\color{black}{an}} arbitrary size WLMMSE MIMO system. The so-obtained PDFs have enabled meaningful performance evaluations of WLMMSE MIMO receivers in terms of the outage probability, the symbol error rate, and the diversity gain, all presented in closed form. Finally, the theoretical findings have been verified through Monte Carlo simulations.

{\color{black} As future work, since the rectilinear and QR modulations considered in this paper are with equal power, which relate to a small portion of practical communication applications, we shall further extend our PDF distribution analysis for transmit signals with different phase offsets and unequal powers. This  is mathematically achievable, because more general OQPSK-type signals, e.g., minimum-shift keying and Gaussian minimum-shift keying, are essentially the sum of several amplitude-modulated OQPSK components. Besides, when the number of transmit antennas $N_t > 3$, we can expect that the analytic PDF of the SINR becomes more complicated than \eqref{Eq:WL_SINR_PDF_Nrx3}. A worthwhile attempt is to investigate the analytic SINR distribution in terms of the cumulative distribution function, rather than the PDF. In this way, the Whittaker-M function might enable an analytic solution, expressed in a finite series \cite{gradshteyn2007table}.
}

\begin{appendices}
	\section{The Gamma Random Variable and Its Distribution}\label{App:GammaRV}
	Let $ \{X_p\} \sim \mathcal{N}(0,\,\delta^{2})$ be a set of i.i.d. Gaussian distributed RVs, where $p=1,2,...,P$. Then, $ Z=c\sum\limits_{p = 1}^{P} {{X_p^2}}, \, \,  \forall c \in \mathbb{R} > 0 $, is a gamma-distributed RV, denoted by $Z \sim {\rm Gamma}(\alpha,\,\beta) $ with a shape parameter $ \alpha = P/2 $ and a scale parameter $\beta = 2c\delta^{2}$. Its PDF is given by
	\begin{equation}\label{Eq:lemma1}
	f(z) = \frac{1}{\Gamma(\alpha)\beta^{\alpha}}z^{\alpha-1}e^{-z/\beta},
	\end{equation}
where $0<z<\infty$, $\alpha>0$, and $\beta>0$.
	
	\section{The Distribution of a Linear Combination of Gamma Random Variables }\label{App:SumofGammaRVwithDiffParas}
	Let $ \{X_i\} \sim {\rm Gamma}(\alpha_i,\,\beta_i), \alpha_i>0, \beta_i>0 $ be a set of i.i.d. gamma distributed RVs with different shape and scale parameters, where $i=1,2,...,P$. Then, the PDF of $ Z=\sum\limits_{i = 1}^{P} {{X_i}} $ is \cite{Mathai1982}
	\begin{align}
	\!\!\!f({z}) &\!=\! \frac{{{z}^{{N} - 1}{e^{ - \frac{{{z}}}{{{\beta _1}}}}}}}{{\prod\limits_{i = 1}^{{P}} {\beta_i^{\alpha _i} \Gamma(N)} }} \nonumber\\
	&\times\!\!\Phi _2^{(P\!-\!1)}\!\!\left(\!\! {{\alpha _2},{\alpha _3},\!...,{\alpha _{P}};{N};\!\left( \frac{1}{\beta_1}\!\!-\!\!\frac{1}{\beta_2}\right){z},\!...,\left( \frac{1}{\beta_1}\!\!-\!\!\frac{1}{\beta_{P}}\!\!\right)\!{z}} \!\!\right)\!\!,\!\!
	\label{Eq:PDF_OF_GAMMA_WITH_DIFF_PARAS}
	\end{align}
	where $N = \sum\limits_{i = 1}^P {{\alpha _i}}$ is the sum of shape parameters of the gamma distributed RVs.
	
\section{The Real-Valued Wishart Distribution}\label{App:realwishart}
	A random matrix of form ${\bf Z}={\bf X}{\bf X}^{\mathcal T}$, where ${\bf X}\triangleq[{\bf x}_1,{\bf x}_2,\dots,{\bf x}_Q] \in \mathbb{R}^{P \times Q}$  and the $P\times 1$ column vectors, ${\bf x}_1,{\bf x}_2,\dots,{\bf x}_Q$, are i.i.d. Gaussian with zero-mean, is called a real-valued Wishart matrix with $Q$ degrees of freedom. It is denoted by ${\bf Z} \sim  {\mathcal{W}}_{P}({\bm \Sigma},Q)$, in which ${\bm \Sigma} \triangleq E[{\bf x}_q{\bf x}_q^{\mathcal{T}}]$ is the covariance matrix of the column vector ${\bf x}_q$, where  $q=1,2,...,Q$ \cite{muirhead2009aspects}. Especially, when $Q\ge P$ and each entry of the matrix ${\bf X}$ is i.i.d. ${\mathcal N}(0,\delta)$, the Wishart matrix ${\bf Z} \sim {\mathcal{W}}_{P}(\delta{
		\bf I}_{P},Q)$ and the PDF of its ordered eigenvalues $\lambda_{1},\lambda_{2},\dots,\lambda_{P}$, where $\lambda_{1}>\lambda_{2}>\dots>\lambda_{P}>0$, is given by \cite[Corollary 3.2.19]{muirhead2009aspects},
	\begin{align}\label{Eq:realwisharteigpdf}
	\!\!\!\!f_{\bf \Lambda}(\lambda_{1},\lambda_{2},\dots,\lambda_{P}) \!&=\! \frac{\pi ^{\frac{P^2}{2}}}{(2\delta)^\frac{PQ}{2}{\Gamma _{P}(\frac{P}{2})\Gamma _{P}(\frac{Q}{2})}} \nonumber \\ &\times \!\!\prod\limits_{p = 1}^{P} {e^{-\frac{\lambda_p}{2\delta}}} (\lambda_p)^{\frac{Q-P-1}{2}}\prod\limits_{p < u}^{P}{( {{\lambda _p} \!-\! {\lambda _u}} )}.
	\end{align}
\section{Ratio Test for ${f_{{N_r} \!\times\! 2}}(\tau_j)$ in \eqref{Eq:WL_SINR_PDF_Nrx2_5}}\label{App:AppendixC}

Let $A(m)$ denote the infinite series in \eqref{Eq:WL_SINR_PDF_Nrx2_5}, that is,
	\begin{align}
	A(m) &= \frac{{(N_r\!-\!\frac{1}{2})}_m\tau_j^m(N_r\!+\!m\!-\!1)!}{{{(N_r)}_m}2^m\rho^m m!} \nonumber \\ &\times
	U\left(N_r\!+\! m, N_r\!+\! m\!+\!\frac{3}{2},\frac{\tau_j \!+\!1}{2\rho}\right).
	\end{align}
	Then, we have
	\begin{align}
	\!\!\lim_{m \to \infty } &\frac{{A(m \!+\! 1)}}{{A(m)}}
	\nonumber \\
	&= \!\frac{\tau_j}{2\rho} \lim_{m \to \infty } \!\frac{U\left( {N_r\!+\!m\!+\!1,N_r\!+\!m\!+\! \frac{5}{2},\frac{\tau_j + 1}{{2\rho}}} \right)}{U\left( {N_r\!+\! m,N_r\!+\!m\!+\! \frac{3}{2},\frac{\tau_j+1}{{2\rho}}} \right)}.
	\label{Eq:WL_SINR_PDF_Nrx2_CONVER_PROOF}
	\end{align}

	According to \cite[Eq. (13.2.40)]{olver2010nist}, the confluent hypergeometric functions of the second kind in \eqref{Eq:WL_SINR_PDF_Nrx2_CONVER_PROOF} can be equivalently transformed to
	\begin{equation}\label{Eq:KummerTransform}
	\begin{split}
	U&\left( {N_r\!+\!m\!+\!1,N_r\!+\!m\!+\! \frac{5}{2},\frac{\tau_j\! +\! 1}{{2\rho}}} \right)\nonumber \\
	&\!=\!\left(\frac{\tau_j\! +\! 1}{{2\rho}}\right)^{-N_r\!-\!m\!-\!\frac{3}{2} }U\left( -\frac{1}{2},-\! N_r-\!m\!-\!\frac{1}{2},\frac{\tau_j\! +\! 1}{{2\rho}}\right),\\
	U&\left( {N_r\!+\! m,N_r\!+\!m\!+\! \frac{3}{2},\frac{\tau_j\!+\!1}{2\rho}} \right)
	\nonumber \\
	&\!=\! \left(\frac{\tau_j\! +\! 1}{{2\rho}}\right)^{-\! N_r-\!m\!-\!\frac{1}{2}}U\left( -\frac{1}{2},-\! N_r-\!m\!+\!\frac{1}{2},\frac{\tau_j\! +\! 1}{{2\rho}}\right).
	\end{split}
	\end{equation}
	{\color{black}{Upon}} taking them back into the RHS of \eqref{Eq:WL_SINR_PDF_Nrx2_CONVER_PROOF}, {\color{black}{we have}}
	\begin{align}\label{Eq:WL_SINR_PDF_Nrx2_CONVER_PROOF2}
	&\lim_{m \to \infty } \frac{{A(m \!+\! 1)}}{{A(m)}}
	\nonumber \\
	&= \left( {\frac{\tau_j}{{2\rho}} } \right){\left( {\frac{\tau_j  \!+\! 1}{{2\rho}}} \right)^{ - 1}}\lim_{m \to \infty } \frac{{U\left( { - \frac{1}{2}, -\! N_r-\!m\!-\!\frac{1}{2},\frac{\tau_j+1}{2\rho}} \right)}}{{U\left( { - \frac{1}{2}, -\! N_r-\!m\!+\!\frac{1}{2},\frac{\tau_j+1}{{2\rho}}} \right)}}\nonumber\\
	&= \frac{\tau_j}{\tau_j+1} \lim_{m \to \infty } \frac{{U\left( { - \frac{1}{2}, -\! N_r-\!m\!-\!\frac{1}{2},\frac{\tau_j+1}{2\rho}} \right)}}{{U\left( { - \frac{1}{2}, -\! N_r-\!m\!+\!\frac{1}{2},\frac{\tau_j+1}{{2\rho}}} \right)}}.
	\end{align}
	Based on the recurrent nature of the confluent hypergeometric function of the second kind \cite[Eqs. (13.3.9), (13.3.10)]{olver2010nist}, the limitation on the RHS of \eqref{Eq:WL_SINR_PDF_Nrx2_CONVER_PROOF2} becomes
	\begin{align}\label{Eq:WL_SINR_PDF_Nrx2_CONVER_PROOF3}
	&\lim_{m \to \infty } \frac{{U\left( { - \frac{1}{2}, -\! N_r-\!m\!-\!\frac{1}{2},\frac{\tau_j+1}{2\rho}} \right)}}{{U\left( - \frac{1}{2}, -\! N_r-\!m\!+\!\frac{1}{2},\frac{\tau_j+1}{2\rho} \right)}}\nonumber\\
	&\,\,\!=\! \lim_{m \to \infty } \frac{N_r+m+1+\frac{\tau_j+1}{2\rho}\frac{U\left(\frac{1}{2},-\!N_r-\!m+\frac{1}{2},\frac{\tau_j+1}{2\rho}\right)}{U\left(\frac{1}{2},-\!N_r-\!m\!-\!\frac{1}{2},\frac{\tau_j+1}{2\rho}\right)}}{N_r+m+1+\big(\frac{\tau_j+1}{2\rho}-\frac{1}{2}\big)\frac{U\left(\frac{1}{2},-\!N_r-\!m+\frac{1}{2},\frac{\tau_j+1}{2\rho}\right)}{U\left(\frac{1}{2},-\!N_r-\!m\!-\!\frac{1}{2},\frac{\tau_j+1}{2\rho}\right)}}.
	\end{align}
	According to \cite[Eqs. (13.5.3), (13.5.4)]{olver2010nist}, $\forall N_r,m \in \mathbb{Z}^{+}$, the term $\frac{U\left(\frac{1}{2},-\!N_r-\!m+\frac{1}{2},\frac{\tau_j+1}{2\rho}\right)}{U\left(\frac{1}{2},-\!N_r-\!m-\frac{1}{2},\frac{\tau_j+1}{2\rho}\right)}$ is a continued fraction that converges to a meromorphic function of $\phi$, denoted by $B(\phi)$, where $\phi \triangleq \frac{\tau_j+1}{2\rho}$, then \eqref{Eq:WL_SINR_PDF_Nrx2_CONVER_PROOF3} becomes
	\begin{align}\label{Eq:WL_SINR_PDF_Nrx2_CONVER_PROOF4}
	\lim_{m \to \infty } &\frac{{U\left( { - \frac{1}{2}, -\! N_r-\!m\!-\!\frac{1}{2},\frac{\tau_j+1}{2\rho}} \right)}}{{U\left( - \frac{1}{2}, -\! N_r-\!m\!+\!\frac{1}{2},\frac{\tau_j+1}{2\rho} \right)}}
	\nonumber \\
	&\!=\! \lim_{m \to \infty } \frac{N_r\!+\!m\!+\!1\!+\!\phi B(\phi)}{N_r\!+\!m\!+\!1\!+\!(\phi \!-\!\frac{1}{2})B(\phi)}\!=\!1.
	\end{align}
	Finally, by taking \eqref{Eq:WL_SINR_PDF_Nrx2_CONVER_PROOF4} into \eqref{Eq:WL_SINR_PDF_Nrx2_CONVER_PROOF2}, we arrive at
	\begin{equation}\label{eq:convspeed}
	\lim_{m \to \infty } \frac{{A(m \!+\! 1)}}{{A(m)}}=\frac{\tau_j}{\tau_j+1}<1,
	\end{equation}
	which completes the proof of the absolute convergence of \eqref{Eq:WL_SINR_PDF_Nrx2_5}.
	
\section{Detailed derivation of the inverse Laplace transformation on $G(s)$}\label{App:invLaplace}
	Since $G(s)$ in \eqref{Eq:finalMGF} can be considered as the bilateral Laplace transform of the PDF $f(\tau_j)$, let $t=-s$, we have
    \begin{align}\label{Eq:finalMGF2}
    G(t) \!=\!
    (1\!+\!2\rho t)^{-\frac{2N_r-N_t+1}{2}} \!-\! {\frac{(N_t\!-\!1)}{2N_r\!-\!N_t}}t(1\!+\!2\rho t)^{-\frac{2N_r-N_t+1}{2}}.
    \end{align}
    Next, according to the table of Laplace transforms in \cite{widder2015laplace}, the inverse bilateral Laplace transform of the first term on the RHS of \eqref{Eq:finalMGF} is
    \begin{align}\label{Eq:finalPDF_P1}
    f_1(z) = \frac{{{{({z})}^{\frac{{2{N_r} - {N_t} - 1}}{2}}}{e^{ - \frac{{{z}}}{{2\rho }}}}}}{{{{(2\rho )}^{\frac{{2{N_r} - {N_t} + 1}}{2}}}\Gamma \left( {\frac{{2{N_r} - {N_t} + 1}}{2}} \right)}},
    \end{align}
    and that of the second item can be obtained as
    \begin{align}\label{Eq:finalPDF_P2}
    f_2(z) &\!=\! {\frac{(N_t\!-\!1)}{2N_r\!-\!N_t}}\left( {\frac{{2{N_r} - {N_t} - 1}}{{2{z}}} \!-\! \frac{1}{{2\rho }}} \right)\nonumber\\
    &\times\frac{{{{({z})}^{\frac{{2{N_r} - {N_t} - 1}}{2}}}{e^{ - \frac{{{z}}}{{2\rho }}}}}}{{{{(2\rho )}^{\frac{{2{N_r} - {N_t} + 1}}{2}}}\Gamma \left( {\frac{{2{N_r} - {N_t} + 1}}{2}} \right)}}.
    \end{align}
    A summation of \eqref{Eq:finalPDF_P1} and \eqref{Eq:finalPDF_P2} gives the approximate PDF $f_a(\tau_j)$ in \eqref{Eq:finalapproxPDF}.

\end{appendices}

\bibliographystyle{IEEEtran}
\bibliography{bibliography}

\end{document}